\documentclass[10pt,a4paper]{article}

\usepackage{amsmath,amsgen}
\usepackage{amstext,amssymb,amsfonts,latexsym}
\usepackage{theorem}
\usepackage{pifont}

 \newcommand{\bs}{\bigskip}
 \newcommand{\ms}{\medskip}
 \newcommand{\n}{\noindent}
 \newcommand{\s}{\smallskip}
 \newcommand{\hs}[1]{\hspace*{ #1 mm}}
 \newcommand{\vs}[1]{\vspace*{ #1 mm}}

 \newcommand{\setempty}{\varnothing}
 \newcommand{\real}{\mathbb{R}}
 \newcommand{\nat}{\mathbb{N}}
 
 \newcommand{\integer}{\mathbb{Z}}
 
 \newcommand{\complex}{\mathbb{C}}

 \newcommand{\co}{\mathrm{co}\mbox{-}}

 \newcommand{\BB}{{\cal B}}
 \newcommand{\CC}{{\cal C}}

 \newcommand{\OO}{{\cal O}}
 
 \newcommand{\TT}{{\cal T}}
 \newcommand{\PP}{{\cal P}}

 \newcommand{\VV}{{\cal V}}

 \newcommand{\reg}{\mathrm{REG}}
 \newcommand{\cfl}{\mathrm{CFL}}
 
 \newcommand{\dcfl}{\mathrm{DCFL}}

\theoremstyle{plain}
\theoremheaderfont{\bfseries}
 {\theorembodyfont{\rmfamily}%
  }
 {\theorembodyfont{\rmfamily} }
 {\theorembodyfont{\rmfamily} }

 \newenvironment{proofof}[1]{\vspace*{5mm} \par \noindent
         {\bf Proof of #1.\hs{2}}}{\hfill$\Box$ \vspace*{3mm}}

 \newenvironment{proof}{\par \noindent
            {\bf Proof. \hs{2}}}{\hfill$\Box$ \vspace*{3mm}}

 \newtheorem{ytheorem}{Theorem}[section]
 \newtheorem{ylemma}[ytheorem]{Lemma}
 \newtheorem{yproposition}[ytheorem]{{\bf Proposition}}

\newcommand{\ignore}[1]{}

\newcommand{\cent}{{|}\!\!\mathrm{c}}
\newcommand{\dollar}{\$}

\newcommand{\trace}{\mathrm{tr}}

\begin{document}

\pagestyle{plain}
\setcounter{page}{1}

\begin{center}
{\Large {\bf One-Way Topological  Automata and the Tantalizing \s\\
Effects of Their Topological Features}}\footnote{An extended abstract appeared in the Proceedings of the 10th International Workshop on Non-Classical Models of Automata and Applications (NCMA 2018), August 21--22, 2018, Ko\v{s}ice, Slovakia, \"{O}sterreichische Computer Gesellschaft (the Austrian Computer Society), pp. 197--214, 2018.} \bs\ms\\

{\sc Tomoyuki Yamakami}\footnote{Affiliation: Faculty of Engineering, University of Fukui, 3-9-1 Bunkyo, Fukui 910-8507,  Japan} \bs\\
\end{center}

\sloppy

\begin{abstract}
We cast new light on the existing models of one-way deterministic topological automata by introducing a fresh but general, convenient model, in which,  as each input symbol is read, an interior system of an automaton, known as a configuration, continues to evolve in a topological space by applying continuous transition operators one by one. The acceptance and rejection of a given input are determined by observing the interior system after the input is completely processed.
Such automata naturally generalize one-way finite automata of various types, including deterministic, probabilistic, quantum, and pushdown automata.
We examine the strengths and weaknesses of the power of this new automata model when recognizing formal languages. We investigate tantalizing effects of various topological features of our topological automata by analyzing their behaviors  when different kinds of topological spaces and continuous maps, which are used respectively as configuration spaces and transition operators, are provided to the automata. Finally, we present goals and directions of future studies on the topological features of topological automata.

\s
\n{\bf Keywords:}
topological automata, topological space, continuous map, compact, discrete topology, Kolmogorov separation axiom, quantum finite automata
\end{abstract}

\section{Prelude: Background and Current Challenges}\label{sec:introduction}

\subsection{A Historical Account of Topological Automata}

In the theory of computation, \emph{finite-state automata} (\emph{finite automata}, or even  \emph{automata}, for short) are one of the simplest and most intuitive mathematical models to describe ``mechanical procedures,'' each of which depicts a finite number of ``operations'' in order to determine the membership of any given input word to a fixed language. Such procedures have clear resemblance to  physical systems that make \emph{discrete time} evolution, contrary to \emph{continuous time} evolution.
Over decades of their study, these machines have found numerous applications in the fields of engineering, physics, biology, and even economy (see, e.g., \cite{HUM01}).
In particular, a  one-way\footnote{Here, we use the term ``1-way'' to exclude the use of $\lambda$-moves, which are particular transitions of the machine with its tape head staying still, where $\lambda$ refers to the empty string. On the contrary, finite automata that make $\lambda$-moves are sometimes called  \emph{1.5-way} finite automata.} (or real-time) finite automaton reads input symbols one by one and then processes them simply by changing a status of the automaton's interior system step by step. This machinery has been used to  model online data processing, in which it receives streamlined input data and processes such data piece by piece by applying operations predetermined for each of the input symbols.

To cope with numerous computational problems, various types of finite automata have been proposed as their appropriate machine models in the past literature. As a few clear examples, deterministic finite automata were modified to probabilistic finite automata \cite{Rab63}, which were further extended to generalized finite automata \cite{Tur69}. Recent models of quantum finite automata \cite{MC00,KW97} have also extended probabilistic finite automata.
In the 1970s, nonetheless, many features of the known 1-way finite automata were generalized into so-called ``topological automata'' (see \cite{EK74} for early expositions and references therein). Here, a \emph{topology} refers to a mathematical concept of dealing with open sets and continuous maps that preserve the openness of point sets.
More general automata were also defined in terms of \emph{category} in, e.g., \cite{Gog72}.
Topological automata embody characteristic features of various types of finite automata, and therefore this fact has helped us take a unified approach toward the study of formal languages and automata theory. The analysis of topological features of the topological automata can guide us to the better understandings of the theory itself.

Back in the 1970s, Brauer (see references in \cite{EK74}) and Ehrig and K\"{u}hnel \cite{EK74} discussed topological automata as a topological generalization of \emph{Mealy machines}, which produce outputs as they read given inputs. In contrast,
following a discussion of Bozapalidis \cite{Boz03} on a generalization of stochastic functions and quantum functions (see also \cite{Yam03}),
Jeandel \cite{Jea07} studied another type of topological automata that behave as ``acceptors'' of inputs.
Jeandel's model naturally generalizes not only \emph{probabilistic finite automata} \cite{Rab63} but also \emph{measure-once quantum finite automata} \cite{MC00}. The main motivation of Jeandel's work was, nonetheless, to study a nondeterministic variant of quantum finite automata and he then used his  topological automata to obtain an upper-bound of the language recognition power of nondeterministic quantum finite automata.
Another difference concerns the types of ``inputs'' fed into topological automata. Ehrig and K\"{u}hnel \cite{EK74} set up a quite general framework to treat inputs taken from arbitrary \emph{compactly generated Hausdorff spaces},  whereas Jeandel \cite{Jea07} used the standard framework based on finite alphabets and languages generated over them. Jeandel further took ``measures'' (which assign real numbers to final configurations) to determine the acceptance or rejection of inputs.
In this work, since we are more concerned with the computational power of topological automata in comparison with the existing finite automata, we wish to make our model as simple and intuitive as possible by introducing, unlike the use of measures,  sets of accepting and rejecting configurations,  into which the machine's interior system finally fall.

Given an input string over a fixed alphabet $\Sigma$, the evolution of an interior status of our topological automaton is described in the form of a series of \emph{configurations}, which constitutes a \emph{computation} of the machine.
A list of transition operators thus serves as a ``program'', which completely dictates the behaviors of the machine on each input. Since arbitrary topological spaces can be used as \emph{configuration spaces}, topological automata are no longer ``finite-state'' machines; however, they evolve sequentially as they read input symbols one by one until they completely read the entire inputs and final configurations are observed once (referred to as an ``observe once'' feature).
Moreover, our topological automata enjoy a ``deterministic'' nature in the sense that which transition operators are applied to the current configurations is completely determined by input symbols alone. This gives rise to a model of \emph{1-way deterministic topological automata}
(or \emph{1dta's}, for short).
Although their tape heads move in one direction from the left to the right, 1dta's turn out to be quite powerful in recognizing formal languages. By extending transition maps to ``multi-valued'' maps, it is possible to consider nondeterministic moves of topological automata \cite{Jea07}.

For other use of topology in computation, refer to, e.g., \cite{Yam95} and references therein.

\subsection{A New Model of Topological Automata}\label{sec:new-model}

All the aforementioned models of topological automata are based only on a relatively small range of appropriately defined topologies, such as compactly generated Hausdorff spaces. We instead wish to study all possible topologies with no initial restrictions other than discrete applications of
transition operators.

This paper thus aims at shedding new light on the basic structures of topological automata and the acting roles of their transition operators that force  configurations to evolve consecutively. For this purpose, we start our study with a suitable abstraction of 1-way finite automata using arbitrary topological spaces for configurations and arbitrary continuous maps for transitions. Such an abstraction serves as a skeleton to construct our topological automata. We call this skeleton an \emph{automata base}. Since the essential behaviors of topological automata are strongly influenced by the choice of their automata bases, we are mostly concerned with the properties of these automata bases.

In general, the choice of topologies significantly affects the computational power of topological automata. As shown later, the \emph{trivial topology}  induces the language family composed only of $\setempty$ and $\Sigma^*$ (for each fixed alphabet $\Sigma$) whereas the \emph{discrete topology} allows topological automata to recognize all languages.
All topologies on a fixed space $V$ form a \emph{complete lattice}; thus, it is possible to classify the topologies according to the endowed power of associated topological automata.

We suggest that a study on topological automata should be focused on achieving the following four key goals.

\begin{enumerate}
  \setlength{\topsep}{-2mm}%
  \setlength{\itemsep}{1mm}%
  \setlength{\parskip}{0cm}%

\item Understand how various choices of topological spaces and continuous maps affect the computational power of underlying  machines by clarifying the strengths and weaknesses of the language recognition power of the machines.

\item Determine what kinds of topological features of topological automata nicely characterize the existing finite automata of various types by examining the descriptive power of such features.

\item Explore different types of topological automata to capture fundamental properties (such as closure properties) of formal languages and the existing finite automata.

\item Find useful applications of topological automata to other fields of science.
\end{enumerate}

\vs{-1}
\paragraph{Organization of the Paper.}
After a brief introduction of topological concepts, we will formulate our fundamental computational model of 1dta's in Section \ref{sec:basic-models}. These 1dta's are naturally induced from automata bases and, in Section \ref{sec:generic-finite-automata}, we will show that the 1dta's have enormous expressive power to describe numerous types of the existing 1-way finite automata. Through Section \ref{sec:basic-properties}, we will discuss basic properties of the 1dta's, including closure properties and the elimination of two endmarkers.
In Section \ref{sec:compact-property}, we will show that unique features of well-known topological concepts, such as  \emph{compactness} and \emph{equicontinuity}, help us characterize 1-way deterministic finite automata (or 1dfa's). In particular, we will lay out a necessary and sufficient condition on a topological space for which its underlying machines are no more powerful than 1dfa's.
Following an exploration of basic properties, we will compare the strengths of different topologies in Section \ref{sec:compare-properties} by measuring how much computational power is endowed to underlying topological automata. In particular, we will discuss the strengths of the trivial topology, the discrete topology, and topologies that violate the \emph{Kolmogorov separation axiom}.
In Section \ref{sec:nondeterminism}, we will consider a nondeterministic variant of our topological automata (called \emph{1nta's}) by introducing \emph{multi-valued} transition operators. It is known that, for weak machine models (such as finite automata), nondeterministic machines can be simulated by deterministic ones at the cost of exponentially more inner states than the nondeterministic ones.
By formalizing this situation, we will argue what kind of topology makes 1nta's simulatable by 1dta's.

We strongly hope that this work reignites a systematic study on the tantalizing effects and features of various topologies used to define topological automata and that, since topological automata can characterize ordinary finite automata of numerous types, this work leads to better understandings of ordinary finite automata as well.

\section{Basics of Topologies and Automata Bases}\label{sec:basic-models}

One-way deterministic topological automata can express the existing one-way finite automata of numerous types.
We begin our study on such powerful automata by describing their basic   framework, which we intend to call an \emph{automata base}, founded solely on topological spaces and continuous maps.
In the subsequent subsections, we will provide a fundamental notion of such automata bases as a preparation to the further exploration of
their properties.

\subsection{Numbers, Sets, and Languages}

Let $\integer$, $\real$,  and $\complex$ respectively indicate the sets of all \emph{integers}, of all \emph{real numbers}, and of all \emph{complex numbers}.
Given a real number $e\geq0$, let $\complex^{\leq e} = \{\alpha\in\complex\mid |\alpha|\leq e\}$ and $\complex^{=e} = \{\alpha\in\complex\mid |\alpha| = e\}$ for simplicity.
We denote by $\nat$ the set of all \emph{natural numbers} (i.e., nonnegative integers) and define $\nat^{+}$ to be $\nat-\{0\}$. For any two integers $m$ and $n$ with $m\leq n$, an \emph{integer interval} $[m,n]_{\integer}$ expresses the set   $\{m,m+1,m+2,\ldots,n\}$ in contrast with a real interval $[\alpha,\beta]$ for two real numbers $\alpha$ and $\beta$ with $\alpha\leq \beta$.
We further abbreviate $[1,n]_{\integer}$ as $[n]$ for each number $n\in\nat^{+}$.

An \emph{alphabet} refers to a nonempty finite set of ``symbols'' or ``letters''.  A  \emph{string} over an alphabet $\Sigma$ is a finite sequence of symbols in $\Sigma$ and the \emph{length} $|x|$ of a string $x$ is the total number of symbols used to form $x$. In particular, the \emph{empty string} is a unique string of length $0$ and is denoted by $\lambda$. Given three strings $x$, $y$, and $z$ over the same alphabet,  when $z=xy$ holds, $x$ is a \emph{prefix} of $z$ and $y$ is a \emph{suffix} of $z$. For each number  $n\in\nat$, $\Sigma^n$ expresses the set of all strings of length exactly $n$; moreover, we set  $\Sigma^* = \bigcup_{n\in\nat}\Sigma^n$ and $\Sigma^+=\Sigma^*-\{\lambda\}$.
Any subset of $\Sigma^*$ is called a \emph{language} over $\Sigma$. In contrast, for each $n\in\nat$, $\Sigma^{\leq n}$ refers to the set $\{x\in\Sigma^*\mid |x|\leq n\}$.
A language is called \emph{unary} (or \emph{tally}) if it is defined over a single-letter alphabet. Given a language $L$ over $\Sigma$, we use the same symbol $L$ to denote its \emph{characteristic function}; that is, for any $x\in\Sigma^*$, $L(x)=1$ if $x\in L$, and $L(x)=0$ otherwise.
For two languages $A$ and $B$ over $\Sigma$, the notation $AB$ denotes the language $\{xy\mid x\in A,y\in B\}$. In particular, when $A$ is a singleton $\{s\}$, we write $sB$ in place of $\{s\}B$; similarly, we write $As$ for $A\{s\}$. With two special symbols $\cent$ and $\dollar$, for example, the set $\Sigma^*\cup \{\cent\}\Sigma^*\{\dollar,\lambda\}$ coincides with $\{x,\cent{x},\cent{x}\dollar\mid x\in\Sigma^*\}$ and it is later denoted by $\Sigma^*_{\cent\dollar}$.
The \emph{reversal} of a string $x=x_1x_2\cdots x_{n-1}x_n$ with $x_i\in\Sigma$ for any $i\in[n]$ is $x_nx_{n-1}\cdots x_2x_1$ and is denoted by $x^{R}$.

Given a set $X$, the notation $\PP(X)$ denotes the \emph{power set} of $X$, i.e., the set of all subsets of $X$, and $\PP(X)^{+}$
expresses $\PP(X)-\{\setempty\}$.

\subsection{Topologies and Related Notions}\label{sec:topology-notion}

Let us briefly review basic terminology in the theory of
\emph{general topology} (or \emph{point-set topology}).
Given a set $V$ of \emph{points}, a \emph{topology} $T_V$ on $V$ is a  collection of subsets of $V$, which are called \emph{open sets}, such that $T_V$ satisfies the following three axioms: (1) $\setempty,V\in T_V$,  (2) any (finite or infinite) union of sets in $T_V$ is also in $T_V$, and (3) any finite intersection of sets in $T_V$ belongs to  $T_V$.
Notice that $T_{V}$ is a subset of $\PP(V)$. With respect to $V$, the \emph{complement} of each open set of $V$ is called a \emph{closed set}. We write $\co T_V$ for the collection of all closed sets.  Moreover, a \emph{clopen set} is a set that is both open and closed. Clearly, $\setempty$ and $V$ are clopen with respect to $V$.
A \emph{neighborhood} of a point $x$ in $V$ is an open set of $V$ that contains $x$. We often write $N_x$ to indicate
such a neighborhood of $x$.

A \emph{topological space} $(V,T_V)$ is a pair of a point set $V$ and its topology $T_V$. When $T_V$ is clear from the context, we often omit ``$T_V$''  and simply call $V$ a topological space. For a practical reason, we implicitly assume that $V\neq\setempty$ throughout this paper. For simplicity, we write $T_{V}^{+}$ for $T_{V}-\{\setempty\}$.
Given two topological spaces $(V_1,T_{V_1})$ and $(V_2,T_{V_2})$, we say that $(V_2,T_{V_2})$ is \emph{finer} than $(V_1,T_{V_1})$ (also $(V_1,T_{V_1})$ is \emph{coarser} than $(V_2,T_{V_2})$) if both $V_1\subseteq V_2$ and $T_{V_1}\subseteq T_{V_2}$ hold.
In such a case, we write $(V_1,T_{V_1})\sqsubseteq (V_2,T_{V_2})$, or simply $T_{V_1}\sqsubseteq T_{V_2}$ when both $V_1$ and $V_2$ are clear from the context.
For a topological space $V$, a \emph{basis} of its topology $T_{V}$ is a collection $B$ of subsets of $V$ such that every open set in $T_V$ is expressed as a union of sets of $B$. In this case, the basis $B$ is said to \emph{induce} the topology $T_V$. Given two topological spaces $V$ and $W$, the \emph{product topology} (or Tychonoff topology) $T_{V\times W}$ on the Cartesian product $V\times W$ is the topology induced by the basis $\{A\times B\mid A\in T_V,B\in T_W\}$.
For any topological space $(V,T_V)$, a \emph{subspace} $(C,T_C)$ is made up of a subset $C$ of $V$ and a \emph{subspace topology} $T_C$ on $C$ induced by $T_V$, which is defined as $T_C =\{A\cap C\mid A\in T_V\}$.
This subspace $(C,T_C)$ is also a topological space.

Take a point set $V$ and consider all possible topologies on $V$. The collection of all topologies on $V$, denoted by $\TT(V)$, forms a \emph{complete lattice} in which the \emph{join} and the \emph{meet} of a collection $A$ of topologies on $V$ respectively correspond to the intersection of all elements in $A$ and the meet of the collection of all topologies on $V$ that contain every element of $A$.

There are two typical topologies on $V$: the \emph{trivial topology} $T_{trivial}(V)=\{\setempty,V\}$ and the \emph{discrete topology} $T_{discrete}(V) = \PP(V)$. Notice that any topology $T_V$ on $V$ is located between $T_{trivial}(V)$ and $T_{discrete}(V)$ in the lattice $\TT(V)$.

Let us consider a map $B$ from a topological space $V$ to another topological space $W$. We write $B:V\to W$. Given any point $v$ in $V$, the notation $B(v)$ denotes the point of $W$ to which $M$ maps $v$. For two maps $A:U\to V$ and $B:V\to W$, $A\circ B$ (or simply $AB$) denotes the \emph{functional composition} of $A$ and $B$, which is defined as $A\circ B(v) = A(B(v))$ for any $v\in U$. A map $B$ on $V$ (i.e., from $V$ to $V$ itself) is said to be \emph{continuous} if, for any $v\in V$ and any neighborhood $N$ of $B(v)$, there exists a neighborhood $N'$ of $v$ satisfying $B(N')\subseteq N$, where $B(N') = \bigcup_{v\in N'}B(v)$; equivalently, for any neighborhood  $N$ in $V$, the \emph{inverse image}  $B^{-1}(N)$ defined as $\{v\in V\mid B(v)\in N\}$ is an open set in $V$.
The notation $C(V)$ denotes the set of all continuous maps on $V$. Notice that $C(V)$ contains the identity function $I$ and
is \emph{closed under functional composition} $\circ$; namely, for any two maps $A,B\in C(V)$, their functional composition $A\circ B$ also belongs to $C(V)$.

\subsection{Automata Bases}\label{sec:automata-base}

In the 1970s, \emph{topological automata} were sought to take inputs from arbitrary topological spaces (e.g., \cite{EK74}), as noted in Section \ref{sec:introduction}.  In this work, however, we wish to limit our interest within \emph{fixed discrete alphabets} because our intention is to compare the language recognition power of topological automata with the existing  finite automata that recognize languages over discrete alphabets.
We strongly believe that such a treatment of discrete inputs provides a bird's-eye view of a topological landscape inside formal languages and automata theory.

To discuss structures of our topological automata, we first introduce a fundamental notion of ``automata base,'' which is a skeleton of various topological automata introduced in Section \ref{sec:generic-finite-automata}.

\s
{\bf Automata Bases.}
A triplet $(\VV,\BB,\OO)$ is called an \emph{automata base} if $\VV$, $\BB$, and $\OO$ are all nonempty sets and satisfy all of the following conditions.
\begin{enumerate}
  \setlength{\topsep}{-2mm}%
  \setlength{\itemsep}{1mm}%
  \setlength{\parskip}{0cm}%

\item $\VV$ is composed of topological spaces $(V,T_V)$ (which are called \emph{configuration spaces}).

\item $\BB$ consists of subsets $F_V$ of $C(V)$ for each space $V$ in $\VV$ such that $F_V$ is closed under functional composition (where all continuous maps in $C(V)$ are particularly called \emph{transition operators}).

\item $\OO$ is a set of \emph{observable pairs} $(E_{acc},E_{rej})$, where $E_{acc}$ and $E_{rej}$ are both clopen\footnote{In this paper, we demand the clopenness of $E_{acc}$ and $E_{rej}$. It is, however, possible to require only the openness.} sets in a certain space $V$ in $\VV$ (where $E_{acc}$ and $E_{rej}$ are respectively called by an \emph{accepting space} and a \emph{rejecting space}).
\end{enumerate}

Notice that $\BB$ is not required to include $C(V)$ for each $V\in\VV$. Hereafter, by identifying $(V,T_V)$ with $V$, we succinctly write ``$V\in\VV$'' in place of ``$(V,T_V)\in \VV$'' as long as its topology $T_V$ is clear from the context.

It is often convenient to deal with a pair $(\VV,\BB)$ excluding $\OO$; therefore, this pair $(\VV,\BB)$ is particularly called a \emph{sub-automata base}.
Given a map $B: V\to V$, we say that $\OO$ is \emph{closed under} $B$ if  $(B(E_1),B(E_2))\in\OO$ holds for any pair $(E_1,E_2)\in\OO$. Given a ``property''\footnote{This informal term ``property'' is  used in a general sense throughout this paper, not limited to ``topological properties,'' which usually means the ``properties invariant under homeomorphisms.''}  $P$ associated with topological spaces, when all topological spaces in $\VV$ satisfy $P$, we succinctly say that $\VV$ \emph{satisfies} $P$.

\section{One-Way Deterministic Topological  Automata}\label{sec:generic-finite-automata}

We formally  describe in Section \ref{sec:basic-model} our machine model of \emph{one-way deterministic topological automata} (or \emph{1dta's}, for short), which are built upon appropriately chosen automata bases. To shed light on the \emph{expressiveness} of our 1dta's, we demonstrate in Section \ref{sec:1dta-characterize} how the existing finite automata of various types can be completely reformulated in terms of our topological automata.

\subsection{Basic Models of $(\VV,\BB,\OO)$-1dta's}\label{sec:basic-model}

Formally, let us introduce our topological automata, each of which reads input symbols one by one taken from a fixed discrete alphabet, modifies its configurations step by step in a deterministic manner, and finally observes the final configurations to determine the acceptance or rejection of the given inputs. The last step of making an observation could be compared to ``measurement'' in  quantum computation. For quantum finite automata, there are usually two types of measurement, known as ``measure-once'' and ``measure-many'' measurements, in use. As a natural analogy, our model may be called ``observe once,'' because we observe the final configuration \emph{once} after a computation terminates,  instead of observing configurations at every step of the topological automaton.

Hereafter, let $(\VV,\BB,\OO)$ denote an arbitrary automata base.
Customarily, we use two endmarkers $\cent$ (left-endmarker) and $\dollar$ (right-endmarker) to surround an input string $x$ as $\cent x\dollar$ to mark the beginning and the ending of the input $x$.
Without any endmarker, for instance, machines must process a given input string with no knowledge of the end of the string.

\s
{\bf Framework of $(\VV,\BB,\OO)$-1dta's.}
Assuming an arbitrary input alphabet $\Sigma$ with $\cent,\dollar\notin \Sigma$, let us define a basic model of our topological automata. An \emph{1-way (observe-once\footnote{It is possible to consider an \emph{observe-many} model of 1dta in which, at each step,  the 1dta checks if the current configuration falls into $E_{acc}\cup E_{rej}$. For a further discussion, refer to Section \ref{sec:discussion}.})  deterministic $(\VV,\BB,\OO)$-topological  automaton} with the endmarkers (succinctly called a \emph{$(\VV,\BB,\OO)$-1dta}) $M$ is a septuplet
$(\Sigma,\{\cent,\dollar\}, V, \{B_{\sigma}\}_{\sigma\in\check{\Sigma}}, v_0, E_{acc},E_{rej})$,
where $\check{\Sigma}=\Sigma\cup\{\cent,\dollar\}$ is an \emph{extended alphabet}, $V$ is a configuration space in $\VV$ with a certain topology $T_{V}$ on $V$, $v_0$ is the \emph{initial configuration} in $V$, $\{B_{\sigma}\}_{\sigma\in\check{\Sigma}}$ consists of transition operators acting on $V$ taken from
a certain fixed subset $F_V$ of $C(V)$ in $\BB$, and
$(E_{acc},E_{rej})$ is an observable pair in $\OO$ for $V$ satisfying
the following \emph{exclusion principle}: $E_{acc}$ and $E_{rej}$ are disjoint (i.e., $E_{acc}\cap E_{rej}=\setempty$).
For convenience, we write $E_{non}$ for $V- (E_{acc}\cup E_{rej})$.
Notice that the use of the two endmarkers helps us avoid an  introduction of a special transition operator associated with the empty string $\lambda$.

Our definition of 1dta's is different from the existing topological automata in the past literature in the following points.
Ehrig and K\"{u}hnel \cite{EK74} took compactly generated Hausdorff spaces in place of our $\Sigma$ and $V$. Jeandel \cite{Jea07} took a metric space for $V$ and also used a measure mapping $V$ to $\real$ instead of our observable pair $(E_{acc},E_{rej})$.
Concerning our transition operators $\{B_{\sigma}\}_{\sigma\in\check{\Sigma}}$,
as another possible formulation, we may be able to use a single map $B:\check{\Sigma}\times V\to V$ as in \cite{EK74}. Nevertheless, they pointed out  as a drawback  that $B$ is \emph{no longer} continuous.

\s
{\bf Configurations and Computation.}
Let $x=x_1 x_2\cdots x_n$ denote any input string of length $n$ in $\Sigma^*$ and   set $\tilde{x} = x_0x_1\cdots x_nx_{n+1}$ to be an \emph{endmarked input string}, including $x_0=\cent$ (left-endmarker) and $x_{n+1}=\dollar$ (right-endmarker). This new string $\tilde{x}$ can be considered as
a string over the extended alphabet $\check{\Sigma}$.

Our 1dta $M$ works as follows. A \emph{configuration} of $M$ on $x$ is a point of $V$. A configuration in $E_{acc}$ (resp., $E_{rej}$) is called an \emph{accepting configuration} (resp., a \emph{rejecting configuration}). Accepting configurations as well as rejecting configurations are collectively called \emph{halting configurations}.
A computation of $M$ on $x$ begins with the initial configuration $v_0\in V$, which is the \emph{$0$th configuration} of $M$ on $x$.
At the $1$st step,
we apply $B_{\cent}$ to $V_0$ and obtain the $1$st configuration $v_1=B_{\cent}(v_0)$.
For any index $i\in[n]$, we assume that $v_i$ is the $i$th configuration of $M$ on $x$.
At Step $i+1$ ($0\leq i \leq n$), the $(i+1)$th configuration $v_{i+1}$ is obtained from $v_i$ by  applying an operator $B_{x_i}$ chosen according to  $x_i$; namely,  $v_{i+1}=B_{x_i}(v_i)$.
For any finite series $\sigma_1,\sigma_2,\ldots,\sigma_{j-1},\sigma_j\in \check{\Sigma}$, we abbreviate the functional composition    $B_{\sigma_{j}}\circ B_{\sigma_{j-1}}\circ \cdots \circ  B_{\sigma_2}\circ B_{\sigma_1}$  as $B_{\sigma_1\sigma_2\cdots \sigma_j}$.
To describe the behavior of $M$, it suffices to consider only
maps of the form  $B_x$ for any string $x\in\Sigma^*_{\cent\dollar}$.
Notice that, since all $B_{\sigma}$'s are in $F_V$ and the closure property of $F_V$ under functional composition, the map    $B_{\sigma_1\sigma_2\cdots \sigma_j}$ also belongs to $F_V$.
Therefore, the inclusion $\{B_x\}_{x\in\Sigma^*_{\cent\dollar}} \subseteq F_V$ follows.
The \emph{final configuration} $v_{n+2}$ is obtained from $v_{n+1}$ as $v_{n+2} = B_{\dollar}(v_{n+1})$ and it coincides with $B_{\cent x\dollar}(v_0)$.
The obtained series $(v_0,v_1,\ldots,v_{n+2})$ of configurations forms a \emph{computation} of $M$ on $x$.
In the case where the 1dta has no endmarker, by contrast, a computation $(v_0,v_1,\ldots,v_{n})$ is simply generated by the equality $v_i= B_{x_i}(v_{i-1})$ for every index $i\in[n]$, and the final configuration $v_{n}$ coincides with $B_x(v_0)$.

\s
{\bf Acceptance and Rejection.}
Finally, we determine whether the 1dta {accepts} or {rejects} each input string by checking whether the final configuration $v_{n+2}$ falls into $E_{acc}$ or   $E_{rej}$, respectively.
To be more precise, we say that $M$ \emph{accepts} (resp., \emph{rejects}) $x$ if $v_{n+2}\in E_{acc}$ (resp., $v_{n+2}\in E_{rej}$).
Since $E_{acc}\cap E_{rej}=\setempty$, $M$ cannot simultaneously accepts and rejects $x$.
We say that $M$ \emph{recognizes} $L$
if, for every string $x\in\Sigma^*$, the following two conditions are met:
(1) if $x\in L$, then $M$ accepts $x$ and (2) if $x\notin L$, then  $M$ rejects $x$.
The notation $L(M)$ indicates the language that is recognized by $M$.
We define $(\VV,\BB,\OO)\mathrm{\mbox{-}1DTA}$ to be the family of all languages, each of which is defined over a certain alphabet $\Sigma$ and is recognized by a certain $(\VV,\BB,\OO)$-1dta working over $\Sigma$.

Two 1dta's $M_1$ and $M_2$ having the common sets $\Sigma$ and $V$ are said to be \emph{(computationally) equivalent} if $L(M_1)=L(M_2)$. Notice that this equivalence relation satisfies basic properties, including reflexivity, symmetry, and transitivity.

\s
For two topological spaces $V_1$ and $V_2$ together with  a map $f:V_1\to V_2$,  $V_1$ is \emph{homeomorphic} to $V_2$ by $f$ if (i) $f$ is a bijection (thus, $f$ is also \emph{invertible}), (ii) $f$ is continuous, and (iii) the inverse map $f^{-1}$ is continuous. This function $f$ is particularly called a \emph{homeomorphism}.
Given two maps $B_1:V_1\to V_1$ and $B_2:V_2\to V_2$, $B_1$ is \emph{homeomorphic} to $B_2$ via $f$ if, for any pair $v,w\in V_1$, $B_1(v)=w$ implies $B_2(f(v)) = f(w)$. Moreover, two pairs $(A_1,B_1)$ and $(A_2,B_2)$ of sets, $(A_1,B_1)$ is \emph{homeomorphic} to $(A_2,B_2)$ via $f$ if both $A_1$ and $B_1$ are respectively homeomorphic to $A_2$ and $B_2$ via $f|A_1$ and $f|B_1$, where $f|E$ is $f$ \emph{restricted to} $E$.
Let $(\VV,\BB,\OO)$ be any automata base.
For each index $i\in\{1,2\}$, let $M_i=(\Sigma,\{\cent,\dollar\},V_i, \{B_{i,\sigma}\}_{\sigma\in\check{\Sigma}}, v_{i,0},E_{i,acc},E_{i,rej})$ denote an arbitrary $(\VV,\BB,\OO)$-1dta. We say that $M_1$ and $M_2$ are \emph{homeomorphic} if there exists a homeomorphism $f:V_1\to V_2$ such that (1)  $f(v_{1,0}) = v_{2,0}$, (2) $V_1$ is homeomorphic to $V_2$ via $f$,  (3) for every symbol $\sigma\in\check{\Sigma}$, $B_{1,\sigma}$ is homeomorphic to $B_{2,\sigma}$ via $f$, and (4)  $(E_{1,acc},E_{1,rej})$ is homeomorphic to $(E_{2,acc},E_{2,rej})$ via $f$.

As shown below, two homeomorphic 1dta's must recognize
exactly the same languages.

\begin{ylemma}\label{homeomorphic-equiv}
Let $(\VV,\BB,\OO)$ be any automata base and let $M_1$ and $M_2$ denote two  $(\VV,\BB,\OO)$-1dta's. If $M_1$ is homeomorphic to $M_2$, then $M_1$ and $M_2$ are computationally equivalent.
\end{ylemma}

\begin{proof}
For each index $i\in\{1,2\}$, let $M_i=(\Sigma,\{\cent,\dollar\},V_i, \{B_{i,\sigma}\}_{\sigma\in\check{\Sigma}}, v_{i,0},E_{i,acc},E_{i,rej})$ be any $(\VV,\BB,\OO)$-1dta. Assume the existence of a homeomorphism $f$ from $M_1$ to $M_2$. We intend to verify that $L(M_1) = L(M_2)$.
Take any input string $x=x_1x_2\cdots x_n$ of length $n$.
It is possible to prove by induction that, for any index $k\in[0,n+1]_{\integer}$ and any configuration $v\in V_1$,  $B_{1,x_0x_1\cdots x_k}(v_{1,0}) = v$ iff $B_{2,x_0x_1\cdots x_k}(f(v_{1,0})) = f(v)$, provided that $x_0=\cent$ and $x_{n+1}=\dollar$.
If $x$ is in $L(M_1)$, then $B_{1,\cent x\dollar}(v_{1,0})=v_{acc}$ for a certain accepting configuration $v_{acc}\in E_{1,acc}$. Let $v = B_{1,\cent x}(v_{1,0})$. Since $B_{1,\dollar}(v) = v_{acc}$, the homeomorphism $f$ yields both $f(v) = B_{2,\cent x}(f(v_{1,0}))$ and
$B_{2,\dollar}(f(v)) = f(v_{acc})$.
Therefore, we obtain $B_{2,\cent x\dollar}(f(v_{1,0})) = B_{2,\dollar}(B_{2,\cent x}(f(v_{1,0}))) = B_{2,\dollar}(f(v)) = f(v_{acc})$.
Since $E_{1,acc}$ is homeomorphic to $E_{2,acc}$ via $f|E_{1,acc}$,  it follows from $v_{acc}\in E_{1,acc}$ that $f(v_{acc})$ falls into $E_{2,acc}$. This leads to the conclusion that $x\in L(M_2)$.

By a similar argument, we can deduce that $x\notin L(M_1)$ implies $x\notin L(M_2)$ using $E_{1,rej}$ and $E_{2,rej}$. Therefore, we establish the equality $L(M_1) = L(M_2)$.
\end{proof}

As a direct consequence of Lemma \ref{homeomorphic-equiv}, we can freely identify all 1dta's that are homeomorphic to each other.

\subsection{Conventional Finite Automata are 1dta's}\label{sec:1dta-characterize}

Our topological-automata framework naturally extends the existing 1-way finite automata of various types. To support this observation, let us demonstrate that  typical models of 1-way finite automata can be nicely fit into our framework. Such a demonstration clearly exemplifies the usefulness of our formulation of topological automata.

As concrete examples, we here consider only the following types of well-known finite automata studied in the past literature. To comply with our setting of 1dta's, all the finite automata discussed below are assumed to equip with the two endmarkers $\cent$ and $\dollar$.

\s
{\bf (i) Deterministic Finite Automata.}
A \emph{one-way deterministic finite automaton} (or a 1dfa, for short) with the two endmarkers $\cent$ and $\dollar$ can be viewed as a special case of $(\VV,\BB,\OO)$-1dta with an initial configuration $v_0=1$, where $\VV$ equals $\{[k]\mid k\in\nat^{+}\}$ with the discrete topology, $\BB$ contains the set of all maps on $[k]$ for each number $k\in\nat^{+}$, and $\OO$ contains all nonempty partitions $(E_{acc},E_{rej})$ of $[k]$ for each $k\in\nat^{+}$.
Languages recognized by 1dfa's are called \emph{regular} and $\reg$ denotes the set of all regular languages.


\s
{\bf (ii) Probabilistic Finite Automata \cite{Rab63}.}
A \emph{stochastic matrix} is a nonnegative-real matrix in which every column\footnote{Unlike the standard definition, in accordance with our topological automata, we apply each stochastic matrix to column vectors \emph{from the left}, not \emph{from the right} as in any early literature.} sums up to exactly $1$. A \emph{one-way probabilistic finite automaton} (or a 1pfa) is a special case of $(\VV,\BB,\OO)$-1dta, where $\VV= \{[0,1]^{k}\mid k\in\nat^{+}\}$ (in which each point of $[0,1]^k$ is seen as a column vector), $\BB$ contains the set of all $k\times k$ stochastic matrices for each $k\in\nat^{+}$,  and $\OO$ is the set of all pairs $(E_{acc},E_{rej})$, each of which consists of all points $v$ whose projections onto the real intervals $[0,1/2-\varepsilon]$ and $[1/2+\varepsilon,1]$ for certain constants $\varepsilon\in(0,1/2]$.
The notation $\mathrm{1BPFA}$ denotes the set of all languages recognized by 1pfa's with bounded-error probability (i.e., the intervals $[0,1/2]$ and $(1/2,1]$). When unbounded-error probability is allowed, 1pfa's with unbounded-error probability recognize exactly \emph{stochastic languages}. We use the notation $\mathrm{SL}$ for the set of all stochastic languages. It is well-known that  $\mathrm{1BPFA} = \reg$ \cite{Rab63} and  $\reg\subsetneqq \mathrm{SL}$ since $L_{<}=\{a^mb^n\mid m,n\in\nat,m<n\}$ is in $\mathrm{SL}-\reg$.


\s
{\bf (iii) Generalized Finite Automata \cite{Tur69}.}
A \emph{one-way generalized finite automaton} (or a 1gfa), which is a generalization of 1pfa, evolves from an initial real column vector by applying a real square matrix as it reads each input symbol until a final row vector is applied to determine the acceptance/rejection of an input. Such a 1gfa can be seen as a $(\VV,\BB,\OO)$-1dta, where $\VV$ consists of $k$-dimensional real vectors, $\BB$ contains the set of all $k\times k$ real matrices $B$ for any index $k\in\nat^{+}$, and $\OO$ is composed of all pairs $(E_{ac},E_{rej})$ with real spaces $E_{acc}$ and $E_{rej}$ spanned by two disjoint sets of basis vectors.


\s
{\bf (iv) Measure-Once Quantum Finite Automata \cite{MC00}.}
A \emph{measure-once 1-way quantum finite automaton} (or an mo-1qfa), which can be viewed as a quantum extension of bounded-error 1pfa, is allowed to measure its inner state only once after reading off all input symbols. Each mo-1qfa can be described as a $(\VV,\BB,\OO)$-1dta in which $\VV$ is a set of spaces $V = (\complex^{= 1})^k$, $\BB$ contains the set of all $k\times k$ unitary matrices for each index $k\in\nat^+$, and $\OO$ contains all pairs $(E_{acc},E_{rej})$ such that $E_{acc}=\{v\in V\mid \|\Pi_{acc}v\|^2_2>1-\varepsilon\}$ and $E_{rej}=\{v\in V\mid \|\Pi_{rej}v\|^2_2>1-\varepsilon\}$ for a constant $\varepsilon\in[0,1)$ for two projections $\Pi_{acc},\Pi_{rej}$ onto subspaces spanned by disjoint sets of basis vectors, where $\|\cdot\|_{2}$ denotes
the $\ell_2$-norm.
We write $\mathrm{MO\mbox{-}1QFA}$ to denote the collection of all languages recognized by mo-1qfa's with bounded-error probability.


\s
{\bf (v) Measure-Many Quantum Finite Automata \cite{KW97}.}
A \emph{measure-many 1-way quantum finite automaton} (or an mm-1qfa) is a variant of mo-1qfa, which makes a measurement every time the mm-1qfa reads an input symbol. Each mm-1qfa can be described as a $(\VV,\BB,\OO)$-1dta when $\VV$ contains all sets $V$ of the form $(\complex^{\leq1})^{k}\otimes [0,1]\otimes [0,1]$ and $\BB$ contains the set of all maps $T$ defined in \cite[Section 3.2]{Yam14} as
\begin{eqnarray*}
\hs{-2} T(v,\gamma_1,\gamma_2)
 = \left(\Pi_{non}Bv, sgn(\gamma_1)\sqrt{\gamma_1^2 + \|\Pi_{acc}Bv\|^2_2}, sgn(\gamma_2)\sqrt{\gamma_2^2 + \|\Pi_{rej}Bv\|^2_2}\right),
\end{eqnarray*}
where $sgn(\gamma)=+1$ if $\gamma\geq 0$ and $-1$ if $\gamma<0$,
for a certain $k\times k$ unitary matrix $B$ and $3$ projections $\Pi_{acc}$, $\Pi_{rej}$, and $\Pi_{non}$ onto the spaces spanned by disjoint sets of basis vectors.
Concerning bounded-error 1qfa's,  we set $E_{acc}=\{(v,\gamma_1,\gamma_2)\in V \mid \gamma_1^2\geq 1-\varepsilon, |v|^2+\gamma_1^2+\gamma_2^2\leq 1 \}$ and $E_{rej}=\{(v,\gamma_1,\gamma_2)\in V \mid \gamma_2^2\geq 1-\varepsilon, |v|^2+\gamma_1^2+\gamma_2^2\leq 1 \}$ for each constant $\varepsilon\in[0,1/2)$.
Let $\OO$ express the set of all such pairs $(E_{acc},E_{rej})$. For basic properties of $T$, refer to \cite[Appendix]{Yam14}.
We write $\mathrm{MM\mbox{-}1QFA}$ to denote the collection of all languages recognized by bounded-error 1qfa's. It is known that $\mathrm{MO\mbox{-}1QFA} \subsetneqq \mathrm{MM\mbox{-}1QFA} \subsetneqq \reg$.


\s
{\bf (vi) Quantum Finite Automata with Mixed States and Superoperators \cite{ABG+06,FOM09,YS11}.} (see also a survey \cite{AY15})
A \emph{one-way quantum finite automaton with mixed states and superoperators} (or simply, a 1qfa) generalizes both mo-1qfa's and mm-1qfa's.
To describe such a 1qfa over an alphabet $\Sigma$ as a $(\VV,\BB,\OO)$-1dta, for certain indices $k,m\in\nat^{+}$, we define $V$ to be the set of $k$ dimensional vectors, let $v_0 = (1,0,\ldots,0)^T$  in $V$, and let $B_{\sigma}(v) = \sum_{j=1}^{m}A_{\sigma,j} v A_{\sigma,j}^{\dagger}$ for a set $\{A_{\sigma,j}\}_{\sigma\in\check{\Sigma},j\in[m]}\subseteq V$ satisfying $\sum_{j=1}^{m}A_{\sigma,j}^{\dagger} A_{\sigma,j} = I$ (the identity matrix). Let $\Pi_{acc}$ and $\Pi_{rej}$ be projections onto the spaces spanned by disjoint sets of $k$-dimensional basis vectors. We further define $E_{acc} = \{v\in V\mid \trace(\Pi_{acc}v)\geq 1-\varepsilon\}$ and $E_{rej} = \{v\in V\mid \trace(\Pi_{rej}v)\geq 1-\varepsilon\}$ for any constant $\varepsilon\in[0,1)$,  where $\trace(D)$ is the \emph{trace} of a square matrix $D$. Let $\VV$, $\BB$, and $\OO$ respectively consist of all such $V$, $\{B_{x}\}_{x\in\Sigma^*_{\cent\dollar}}$, and $(E_{acc},E_{rej})$.
Each 1qfa is thus expressed as one of the above $(\VV,\BB,\OO)$-1dta's. By $\mathrm{1QFA}$, we indicate the family of all languages recognized by bounded-error 1qfa's. We then obtain $\mathrm{1QFA} = \reg$.


\s
{\bf (vii) Deterministic Pushdown Automata.}
A \emph{one-way deterministic pushdown automaton} (or a 1dpda) $M$ can be seen as a $(\VV,\BB,\OO)$-1dta when $(\VV,\BB,\OO)$ satisfies the following properties. Let $\VV=\{[k]\times \bot\Gamma^*\mid k\in\nat^{+}, \Gamma\,\text{: alphabet}\}$, where $\bot$ is a distinguished bottom marker not in $\Gamma$. For each $k\in\nat^{+}$, $\BB$ contains the set of all maps of the form $B(q,\bot z) = (\mu_1(q,\bot{z}),\mu_2(q,\bot{z}))$ for 2 functions $\mu_1:[k]\times \bot\Gamma^*\to [k]$ and $\mu_2:[k]\times \bot\Gamma^*\to \bot\Gamma^*$, where $z\in\Gamma^*$.
Intuitively, a single application of $B$ represents a series of moves in which $M$ reads one symbol and then makes a single non-$\lambda$-move followed by a certain number of $\lambda$-moves. Let $\OO$ consist of all pairs $(E_{acc},E_{rej})$ with $E_{acc}=  Q_1\times \bot\Gamma^*$ and $E_{rej}= Q_2\times \bot\Gamma^*$, where $\{Q_1,Q_2\}$ is a partition of $[k]$.
We write $\dcfl$ for the class of all languages recognized by 1dpda's. Well known relations include $\reg\subsetneqq \cfl$.

\section{Basic Properties of $(\VV,\BB,\OO)$-1dta's}\label{sec:basic-properties}

For a given automata base $(\VV,\BB,\OO)$, we have formulated the computational model of $(\VV,\BB,\OO)$-1dta's in Section \ref{sec:basic-model} and we have shown in Section \ref{sec:1dta-characterize} that this model has an ability to characterize the existing finite automata of various types.
Here, we plan to explore basic properties of those $(\VV,\BB,\OO)$-1dta's and their associated language family $(\VV,\BB,\OO)\mbox{-}\mathrm{1DTA}$.

\subsection{Elimination of Endmarkers: Markless 1dta's}\label{sec:elimination}

Although the two endmarkers $\cent$ and $\dollar$ play important roles in signaling the beginning and the ending of each input, in many cases, it is possible to eliminate them from a 1dta $M$ without changing its recognized language $L(M)$. A simple way to eliminate the left-endmarker $\cent$ is to modify the initial configuration, say, $v_0$ of $M$ to a new initial configuration $B_{\cent}(v_0)$ using a map $B_{\cent}$ of $M$. Even if we stick to the same $v_0$ instead of introducing $B_{\cent}(v_0)$, a slight modification of all maps $B_{\sigma}$ of $M$ can provide the same effect, as shown in Lemma \ref{elimination-marker}.

We say that a set $\BB$ of families of maps is \emph{continuously invertible} if, for any element $F$ of $\BB$, every map $B$ in $F$ is invertible and its inverse $B^{-1}$ is also in $F$ and continuous.

\begin{ylemma}\label{elimination-marker}
Let $(\VV,\BB,\OO)$ be any automata base and assume that $\BB$ is continuously invertible. For every $(\VV,\BB,\OO)$-1dta $M=(\Sigma,\{\cent,\dollar\},V,\{B_{\sigma}\}_{\sigma\in\check{\Sigma}}, v_0,E_{acc},E_{rej})$, there exists its equivalent $(\VV,\BB,\OO)$-1dta $N$ with the same $\Sigma$, $V$, $v_0$, $E_{acc}$, and $E_{rej}$ but no left-endmarker $\cent$.
\end{ylemma}

\begin{proof}
Let $M = (\Sigma,\{\cent,\dollar\},V, \{B_{\sigma}\}_{\sigma\in\check{\Sigma}}, v_0,E_{acc},E_{rej})$ be any given $(\VV,\BB,\OO)$-1dta. We define a new set  $\{B'_{\sigma}\}_{\sigma\in\check{\Sigma}}$ of maps as follows. For each symbol $\sigma\in\Sigma$, we define $B'_{\sigma}=B_{\cent}^{-1}B_{\sigma}B_{\cent}$  and we set $B'_{\dollar} = B_{\dollar}B_{\cent}$. The desired 1dta $N$ reads an input of the form $x\dollar$ and behaves exactly as $M$ by substituting $B'_{\sigma}$ for $B_{\sigma}$. It thus follows that $L(M) = L(N)$.
\end{proof}

We can eliminate $\dollar$ as well by slightly changing observable pairs of $M$ as stated in Lemma \ref{right-endmarker}.
A set $\OO$ of observable pairs is said to be \emph{closed under inverse images of maps with respect to $\BB$} if, for any element $F\in\BB$, for any operator $B\in F$, and for any pair $(E_1,E_2)\in\OO$, the pair $(B^{-1}(E_1),B^{-1}(E_2))$ also belongs to $\OO$, where $B^{-1}(A)$ denotes the inverse image of $A$ (i.e., $\{v\in V\mid B(v)\in A\}$).
We remark that this notation $B^{-1}$ will be used even if $B$ itself is \emph{not} invertible.

\begin{ylemma}\label{right-endmarker}
Let $(\VV,\BB,\OO)$ be any automata base. Assume that
$\OO$ is closed under inverse images of maps with respect to $\BB$.  For every $(\VV,\BB,\OO)$-1dta $M = (\Sigma,\{\cent,\dollar\},V, \{B_{\sigma}\}_{\sigma\in\check{\Sigma}}, v_0,E_{acc},E_{rej})$, there exists its equivalent $(\VV,\BB,\OO)$-1dta $N$ with the same $\Sigma$, $V$, $B_{\sigma}$, and $v_0$ but no right-endmarker $\dollar$.
\end{ylemma}

\begin{proof}
Let $M$ be any $(\VV,\BB,\OO)$-1dta $M$ in the premise of the lemma. We define a new observable pair $(E'_{acc},E'_{rej})$ of $N$ by setting  $E'_{acc}=\{v\in V\mid B_{\dollar}(v)\in E_{acc}\}$ and $E'_{rej}=\{v\in V\mid B_{\dollar}(v)\in E_{rej}\}$. Clearly, $E'_{acc}$ and $E'_{rej}$ are disjoint because so are $E_{acc}$ and $E_{rej}$. Note that $E'_{acc}$ and $E'_{rej}$ are written as  $B^{-1}_{\dollar}(E_{acc}) = \{B_{\dollar}^{-1}(v)\mid v\in E_{acc}\}$ and $B^{-1}_{\dollar}(E_{rej}) = \{B_{\dollar}^{-1}(v)\mid v\in E_{rej}\}$, respectively.
Since $\OO$ is closed under inverse images of maps in $\BB$,  it follows that $(B^{-1}_{\dollar}(E_1),B^{-1}_{\dollar}(E_2))\in \OO$ for any $(E_1,E_2)\in\OO$.  In particular, since $(E_{acc},E_{rej})\in\OO$, we conclude that $(E'_{acc},E'_{rej})\in \OO$. It is easy to show that $N$ correctly simulates $M$ on all inputs.
\end{proof}

Lemmas \ref{elimination-marker}--\ref{right-endmarker} seem to place a heavy restriction on an underlying automata base $(\VV,\BB,\OO)$. This situation makes the  endmarker elimination so costly. In certain cases, however, the 1dta model with no use of endmarkers, dubbed as  \emph{markless  1dta's}, has a clear advantage in considering the effects of topological features of 1dta's. We will see such a case in Section \ref{sec:compare-properties}.

\subsection{Closure Properties}

Let us discuss \emph{closure properties} of a language family $(\VV,\BB,\OO)$-1DTA induced from an automata base $(\VV,\BB,\OO)$.
We start with \emph{inverse homomorphisms} whose closure property turns out to be met by every family $(\VV,\BB,\OO)\mbox{-}\mathrm{1DTA}$. Given two alphabets $\Sigma$ and $\Gamma$, a \emph{homomorphism} $h$ is a function from $\Sigma$ to $\Gamma^*$ and it is extended to the domain $\Sigma^*$ by setting $h(\lambda)=\lambda$ and $h(xa)= h(x)h(a)$ for any $x\in\Sigma^*$ and any $a\in\Sigma$. We say that a language family $\CC$ is \emph{closed under inverse homomorphism} if, for any language $L$ in $\CC$ and any homomorphism $h$, the inverse image $h^{-1}(L)$ ($=\{x\in\Sigma^* \mid h(x)\in L\}$) also belongs to $\CC$.

\begin{ylemma}
For any automata base $(\VV,\BB,\OO)$, $(\VV,\BB,\OO)\mbox{-}\mathrm{1DTA}$ is closed under inverse homomorphism.
\end{ylemma}

\begin{proof}
Let $\Sigma$ and $\Gamma$ denote two alphabets and consider a homomorphism $h:\Sigma\to\Gamma^*$ and its extension to the domain $\Sigma^*$.
For a given automata base  $(\VV,\BB,\OO)$ and a language $L$, assume that $L$ belongs to $(\VV,\BB,\OO)$-1DTA.
We then take a $(\VV,\BB,\OO)$-1dta $M = (\Gamma,\{\cent,\dollar\}, V,\{B_{a}\}_{a\in\check{\Gamma}}, v_0,E_{acc},E_{rej})$ that recognizes $L$.

Let us define a new $(\VV,\BB,\OO)$-1dta $N = (\Sigma,\{\cent,\dollar\}, V,\{B'_{\sigma}\}_{\sigma\in\check{\Sigma}}, v_0,E_{acc},E_{rej})$ for  $h^{-1}(L)$.
Initially, we set $B'_{\cent}=B_{\cent}$, $B'_{\dollar} = B_{\dollar}$, and  $B'_{\sigma} = B_{h(\sigma)}$ for each symbol $\sigma\in\Sigma$. For convenience, we also set $B'_{\lambda} = B_{\lambda}$ since $h(\lambda)=\lambda$.
By induction, we can prove that $B'_{x}=B_{h(x)}$ for every string $x\in\Sigma^*$. It then follows that, for each  input $x\in\Sigma^*$, $B'_{\cent x\dollar}(v_0) = B'_{\dollar}(B'_{x}(B'_{\cent}(v_0))) = B_{\dollar}(B_{h(x)}(B_{\cent}(v_0))) = B_{\cent h(x)\dollar}(v_0)$. We thus conclude that $x\in L(N)$ iff $h(x)\in L$. From this equivalence,  $L(N) = h^{-1}(L)$ follows. Therefore, $h^{-1}(L)$ is also in $(\VV,\BB,\OO)$-1DTA.
\end{proof}

Next, we consider other fundamental closure properties: \emph{Boolean closures}.  To state our result on Boolean closures in Lemma \ref{closure-basics}, we need new terminology. Let $(\VV,\BB,\OO)$ be any automata base.
We say that $\OO$ is \emph{symmetric} if, for any pair $(A,B)\in\OO$, $(B,A)$  also belongs to $\OO$.
We consider the \emph{product} $(V,T_V)$ of two topological spaces $(V_1,T_{V_1})$  and $(V_2,T_{V_2})$ by setting $V=V_1\times V_2$ and by taking the associated product topology $T_{V} = T_{V_1\times V_2}$.
Given two maps $B_1:V_1\to V_1$ and $B_2:V_2\to V_2$, the notation $B_1\times B_2$ denotes the map $g:  V_1\times V_2\to V_1\times V_2$ defined by $g(x,y)= (B_1(x),B_2(y))$ for any $(x,y)\in V_1\times V_2$.
Note that $B_1\times B_2$ is continuous (with respect to the product topology $T_{V_1\times V_2}$) whenever $B_1$ and $B_2$ are both continuous with respect to $T_{V_1}$ and $T_{V_2}$, respectively.
A sub-automata base $(\VV,\BB)$ is said to be \emph{closed under product} if, for any $V_1,V_2\in \VV$ and any $F_1,F_2\in\BB$ with $F_1\subseteq C(V_1)$ and $F_2\subseteq C(V_2)$, there exist a topological space   $V\in\VV$ and a subset $F_V$ of $C(V)$ in $\BB$ such that
(i) $V_1\times V_2$ is homeomorphic to $V$ and (ii) every set in $\{B_1\times B_2\mid B_1\in F_1,B_2\in F_2\}$ is homeomorphic to a certain element in $F$.
Furthermore, we say that $(\VV,\OO)$ is \emph{closed under accept-union product} if, for any $V_1,V_2\in\VV$ and any $(E_{1,acc},E_{1,rej}),(E_{2,acc},E_{2,rej})\in\OO$, letting $E'_{acc} = (E_{1,acc}\times V_2) \cup (V_1\times E_{2,acc})$ and $E'_{rej} = E_{1,rej}\times E_{2,rej}$, the pair $(E'_{acc},E'_{rej})$ is also homeomorphic to a certain pair in $\OO$. Similarly, we can define the notion of the \emph{closure under reject-union product} by swapping the roles of two subscripts ``acc'' and ``rej'' in the above definition.

\begin{ylemma}\label{closure-basics}
Let $(\VV,\BB,\OO)$ be any automata base.
\begin{enumerate}
  \setlength{\topsep}{-2mm}%
  \setlength{\itemsep}{1mm}%
  \setlength{\parskip}{0cm}%

\item If $\OO$ is symmetric, then $(\VV,\BB,\OO)\mbox{-}\mathrm{1DTA}$ is closed under  complementation.

\item If $(\VV,\BB)$ is closed under product and $(\VV,\OO)$ is closed under accept-union product, then $(\VV,\BB,\OO)\mbox{-}\mathrm{1DTA}$ is closed under union.

\item If $(\VV,\BB)$ is closed under product and $(\VV,\OO)$ is closed under reject-union product, then $(\VV,\BB,\OO)\mbox{-}\mathrm{1DTA}$ is closed under intersection.
\end{enumerate}
\end{ylemma}

We remark that the assumptions of Lemma \ref{closure-basics}(2)--(3) are necessary because, for the automata base $(\VV,\BB,\OO)$ used in Section \ref{sec:1dta-characterize} to define 1dpda's, its sub-automata base $(\VV,\BB)$ is not closed under product. This reflects the fact that the language family $\dcfl$ is  closed under neither union nor intersection.
Similarly, $\mathrm{MM\mbox{-}1QFA}$
is not closed under union \cite{AKV01}.

\begin{proofof}{Lemma \ref{closure-basics}}
(1) The closure property of $(\VV,\BB,\OO)\mbox{-}\mathrm{1DTA}$ under complementation can be obtained simply by exchanging between $E_{acc}$ and $E_{rej}$ since $\OO$ is symmetric.

(2) For each index $i\in\{1,2\}$, we take a language $L_i$ over $\Sigma$ recognized by a certain $(\VV,\BB,\OO)$-1dta $M_i = (\Sigma,\{\cent,\dollar\},V_i,\{B_{i,\sigma}\}_{\sigma\in\check{\Sigma}}, v_{i,0},E_{i,acc},E_{i,rej})$.
Note that $V_i\in\VV$ and $B_{i,\sigma}\in F_i$ for a certain subset $F_i$ of $C(V_i)$ in $\BB$. Consider the union $L=L_1\cup L_2$. For two pairs $(V_1,B_{1,\sigma})$ and $(V_2,B_{2,\sigma})$, consider another pair  $(V,B_{\sigma})$ defined by $V=V_1\times V_2$ and $B_{\sigma}=B_{1,\sigma}\times B_{2,\sigma}$.
Moreover, we set $v_0=(v_{1,0},v_{2,0})$, $E_{acc}=(V_1\times E_{2,acc})\cup (E_{1,acc}\times V_2)$, and $E_{rej}=E_{1,rej}\times E_{2,rej}$.
For any prefix $z$ of $x\dollar$, we obtain $B_{\cent z}(v_0) = (B_{1,\cent z}(v_{1,0}),B_{2,\cent z}(v_{2,0}))$. It thus follows that (i) $B_{\cent x\dollar}(v_0) \in E_{acc}$ iff either $B_{1,\cent x\dollar}(v_{1,0})\in E_{1,acc}$ or $B_{2,\cent x\dollar}(v_{2,0})\in E_{2,acc}$ and (ii) $B_{\cent x\dollar}(v_0) \in E_{rej}$ iff both $B_{1,\cent x\dollar}(v_{1,0})\in E_{1,rej}$ and $B_{2,\cent x\dollar}(v_{2,0})\in E_{2,rej}$.
By the closure property of $(\VV,\BB)$ under product, there exist appropriate elements $\tilde{V}\in \VV$,  $\tilde{F}\in\BB$, $\{B_{\sigma}\}_{\sigma\in\check{\Sigma}}\subseteq \tilde{F}$,  $\tilde{v}_0\in\tilde{V}$, and $(\tilde{E}_{acc},\tilde{E}_{rej})\in \OO$ such that $\tilde{V}$, $\{\tilde{v}_0\}$, $\tilde{B}_{\sigma}$, $\tilde{E}_{acc}$, and $\tilde{E}_{rej}$ are homeomorphic to $V$, $\{v_0\}$,  $B_{\sigma}$, $E_{acc}$, and $E_{rej}$, respectively.
Therefore, it suffices to define the desired machine $N$ as  $(\Sigma,\{\cent,\dollar\},\tilde{V}, \{\tilde{B}_{\sigma}\}_{\sigma\in\check{\Sigma}}, \tilde{v}_0, \tilde{E}_{acc},\tilde{E}_{rej})$.

(3) The proof is similar to (1) in principle, but we need to exchange the roles of ``acc'' and ``rej''.
\end{proofof}

\subsection{Finite Topologies and Regularity}\label{sec:finite-topology}

We briefly discuss a topology composed only of a \emph{finite} number of open sets. We succinctly call such a topology a \emph{finite topology}.
As shown in Section \ref{sec:1dta-characterize}, any 1dfa can be simulated by a certain 1dta with a discrete finite topology. Conversely, we argue in Theorem  \ref{finite-topology} that no finite topology can endow topological automata with more recognition power than 1dfa's.

Two points $x$ and $y$ of a topological space $(V,T_V)$ are said to be \emph{topologically distinguishable} if there exists an open set $P\in T_V$ such that either (i) $x\in P$ and $y\notin P$ or (ii) $x\notin P$ and $y\in P$. Otherwise, they are \emph{topologically indistinguishable}.

\begin{ytheorem}\label{finite-topology}
For any automata base $(\VV,\BB,\OO)$ with finite topologies, it follows that $(\VV,\BB,\OO)\mbox{-}\mathrm{1DTA}\subseteq \reg$.
\end{ytheorem}

\begin{proof}
Let $(\VV,\BB,\OO)$ be any automata base with finite topologies and
consider an arbitrary $(\VV,\BB,\OO)$-1dta  $M = (\Sigma,\{\cent,\dollar\},V, \{B_{\sigma}\}_{\sigma\in\check{\Sigma}}, v_0,E_{acc},E_{rej})$. To show the theorem, we intend to convert $M$ into its equivalent 1dfa $N$. For any two points $v,w\in V$, we write $v\equiv w$ if $v$ and $w$ are topologically indistinguishable.

We claim that this binary relation $\equiv$ is an equivalence relation of finite index. This can be shown as follows. For any point $v\in V$, we define $T(v)$ to be the set of all open sets in $T_V$ that contain $v$. It thus follows that, for any $u,v\in V$, $u\equiv v$ iff $T(u)=T(v)$. This implies that $\equiv$ is an equivalence relation.
Let us consider the set $V\!/{\equiv}$ of all equivalence classes. Since $T_V$ is a finite topology, there are only finitely many different  $T(u)$'s; thus, $|V\!/{\equiv}|$ must be finite.

For the subsequent argument, we set $m = |V\!/{\equiv}|$.
We then choose $m$ points $v_0,v_1,\ldots,v_{m-1}\in V$ satisfying $T(v_i)\neq T(v_j)$ for any distinct pair $i,j\in[0,m-1]_{\integer}$.  We then define the desired 1dfa $N=(Q,\Sigma,\{\cent,\dollar\},\delta,v_0,Q_{acc},Q_{rej})$ as follows.  Let $Q=\{v_0,v_1,\ldots,v_{m-1}\}$ and define two subsets $Q_{acc}=\{v_i\mid i\in[m], T(v_i)\cap E_{acc}\neq\setempty\}$ and $Q_{rej}=\{v_i\mid i\in[m], T(v_i)\cap E_{rej}\neq\setempty\}$.
The transition function $\delta:Q\times\check{\Sigma}\to Q$ is defined as follows: for any pair $i,j\in[m]$, $\delta(v_i,\sigma)=v_j$ iff there are points (not necessarily in $Q$) $w_i,w_j\in V$ such that  $T(v_i)=T(w_i)$, $T(v_j)=T(w_j)$, and $B_{\sigma}(w_i)=w_j$. Let $\delta^*(q,w)$ denote an inner state obtained just after reading $w$ when starting in state $q$. Concerning this $\delta^*$, for any prefix $z$ of $x\dollar$, it follows that $B_{\cent z}(v_0)\in E_{acc}$ iff $\delta^*(v_0,\cent z)\in Q_{acc}$.

Next, we wish to claim that, for any $v,w\in V$, $T(v)=T(w)$ implies $T(B_{\sigma}(v))=T(B_{\sigma}(w))$. To lead to a contradiction, we assume that $T(B_{\sigma}(v))\neq T(B_{\sigma}(w))$. Take a neighborhood $P$ of $B_{\sigma}(v)$ satisfying $B_{\sigma}(w)\notin P$. In the case where a neighborhood $P$ of $B_{\sigma}(w)$ satisfies $B_{\sigma}(v)\notin P$ instead, we should swap the role of $v$ and $w$. Since $B_{\sigma}$ is continuous, we can take another neighborhood $P'$ of $v$ for which $B_{\sigma}(P')\subseteq P$. By the equality $T(v)=T(w)$, $w$ belongs to $P'$. This implies that $B_{\sigma}(w)\in B_{\sigma}(P')$, a contradiction against $B_{\sigma}(w)\notin P$. Thus, the claim should be true.

Finally, we remark that there is no index $i\in[0,m-1]_{\integer}$ such that $T(v_i)\cap E_{acc}\neq\setempty$ and $T(v_i)\cap E_{rej}\neq\setempty$. This is  because, otherwise, there are two distinct points $w_1,w_2\in T(v_i)$ satisfying that $w_1\in E_{acc}$ and $w_2\in E_{rej}$, and thus $w_1\not\equiv w_2$ follows, a contradiction against the choice of $w_1$ and $w_2$.

From the aforementioned properties, we conclude that $N$ simulates $M$ on every input; hence, $L(M)=L(N)$ follows. Therefore, we obtain  $(\VV,\BB,\OO)\mbox{-}\mathrm{1DTA}\subseteq \reg$.
\end{proof}

\subsection{Computational Power Endowed by the Trivial and the Discrete Topologies}

We briefly discuss the language recognition power endowed to 1dta's by the \emph{trivial topology} as well as the \emph{discrete topology}, because all other topologies are located between these two topologies, as discussed in Section \ref{sec:topology-notion}.
In fact, while the trivial topology makes 1dta's recognize only ``trivial'' languages, the discrete topology makes 1dta's powerful enough to recognize all languages. This latter fact, in particular, assures us to be able to characterize any language family by an appropriate choice of topologies for 1dta's, and this further helps us compare the computational strengths of (properties of) topologies.

\begin{yproposition}\label{trivial-topology}
Let $(\VV,\BB,\OO)$ be an automata base with the trivial topology $T_{trivial}(V)$ for every configuration space $V\in\VV$. For any $(\VV,\BB,\OO)$-1dta $M$ with an  alphabet $\Sigma$, $L(M)$ is either $\setempty$ or $\Sigma^*$.
\end{yproposition}

\begin{proof}
Given an automata base $(\VV,\BB,\OO)$ in the lemma, let us consider any $(\VV,\BB,\OO)$-1dta $M=(\Sigma,\{\cent,\dollar\},V,\{B_{\sigma}\}_{\sigma\in\check{\Sigma}}, v_0,E_{acc},E_{rej})$. Since $E_{acc}$ is clopen with respect to $T_{trivial}(V)$, it must be either $\setempty$ or $V$. The same holds for $E_{rej}$. Hence, $M$ either accepts all strings or rejects all strings. From this consequence, we conclude that $L(M)$ is either $\Sigma^*$ or $\setempty$.
\end{proof}

The trivial topology provides little power to 1dta's. In contrast, the discrete topology gives underlying automata enormous computational power so that they can recognize all possible languages.

\begin{yproposition}\label{all-languages}
There is an automata base $(\VV,\BB,\OO)$ with the discrete topology for  each  $V\in\VV$ such that, for any language $L$, there is a $(\VV,\BB,\OO)$-1dta that recognizes $L$. This is true for the 1dta model with or without endmarkers.
\end{yproposition}

\begin{proof}
Let $\VV$ be composed of all languages $\Sigma^*$ for any alphabet $\Sigma$ with the discrete topology on $\Sigma^*$. Moreover, let $\BB$ be composed of all sets $C(V)$ for topological spaces $V$ in $\VV$. Finally, we define $\OO$ as $\{(L,\Sigma^*-L),(\{s_0\},\{s_1\}) \mid L\subseteq\Sigma^*,s_0,s_1\in\Sigma^{+},s_0\neq s_1,\Sigma\text{: alphabet}\}$. Let $L$ be any language over an alphabet $\Sigma$ and set $V=\Sigma^*$.
Since $T_V$ is the discrete topology, clearly $\{s_0\},\{s_1\}\in T_V$ and both $L$ and $\Sigma^*-L$ belong to $T_V$. Therefore, $\OO$ is a set of \emph{valid}  observable pairs.

We want to construct a $(\VV,\BB,\OO)$-1dta $M$ that recognizes $L$. Firstly, let us consider the case where $M$ uses no endmarker. We set $V=\Sigma^*$, $v_0=\lambda$, $B_{\sigma}(v) =v\sigma$, $E_{acc}=L$, and $E_{rej}=\Sigma^*-L$, where $v\sigma$ is the concatenation of $v$ and $\sigma$. These definitions imply that, for any $x\in\Sigma^*$, $B_{x}(v_0)=x\in L$ iff $x\in E_{acc}$. Finally, we define $M$ to be $(\Sigma,V,\{B_{\sigma}\}_{\sigma\in\Sigma}, v_0,E_{acc},E_{rej})$. The construction of $M$ implies that $L=L(M)$.

Next, we want to deal with the case where $M$ uses the two endmarkers.  We fix two distinguished distinct points  $s_0,s_1\in\Sigma^+$.
We define $M = (\Sigma,\{\cent,\dollar\}, V, \{B_{\sigma}\}_{\sigma\in\Sigma}, v_0,E_{acc},E_{rej})$ as follows. Define $V=\Sigma^*$, $v_0=\lambda$, $B_{\cent}=I$, $B_{\sigma}(v)=v\sigma$, $B_{\dollar}(v)=s_{L(v)}$, $E_{acc}=\{s_1\}$, and $E_{rej}=\{s_0\}$, where $L(\cdot)$ denotes the characteristic function of $L$.
It thus follows that, for each $x\in\Sigma^*$, $B_{\cent x}(v_0)=x$ and $B_{\cent x\dollar}(v_0) = B_{\dollar}(x)= s_{L(x)}$. As a result, $x\in L$ implies $B_{\cent x\dollar}(v_0)\in E_{acc}$, and $x\notin L$ implies $B_{\cent x\dollar}(v_0)\in E_{rej}$. We thus conclude that $L=L(M)$.
\end{proof}

Since any language can be expressed in terms of topologies by Proposition \ref{all-languages}, the scrupulous study of topologies would have the significant impact on promoting our understanding of formal languages and ordinary finite automata.

\section{Compactness, Equicontinuity, and Regularity}\label{sec:compact-property}

In general topology, the notion of \emph{compactness} for topological spaces plays an important role. This notion also makes a significant effect on the computational complexity of 1dta's.
For his topological automaton $M$ with a metric space $V$ and a topological space $\BB$ made of continuous maps, Jeandel claimed in \cite[Theorem 3]{Jea07} that the compactness of $V$ and $\BB$ yields the regularity of the language $L(M)$. In contrast, since our topological automata use arbitrary topologies, not limited to metric spaces, we can provide a much more general assertion, which gives a necessary and sufficient condition for the regularity of languages.

To explain our assertion (Theorem \ref{iff-cond-uniform}), we need a few more terminology.
With an appropriate index set $J$, a collection $\{W_i\}_{i\in J}$ of open subsets of $V$ is called a \emph{covering} if $V\subseteq \bigcup_{i\in J} W_i$. A \emph{subcovering} of $\{W_i\}_{i\in J}$ is any subset of $\{W_i\}_{i\in J}$ that is a covering itself.
A subcovering $\{W_i\}_{i\in K}$ with $K\subseteq J$ is said to be \emph{finite} if the index set $K$ is finite. A topological space $(V,T_V)$ is called \emph{compact} if every covering of $V$ has a finite subcovering.


A \emph{uniform structure} on $V$ is a collection $\Phi$ of binary relations on $V$ (equivalently, subsets of $V\times V$) satisfying that (i) all elements of $\Phi$ are reflexive, (ii) $\Phi$ is closed under union with an arbitrary binary relation on $V$, (iii) $\Phi$ is closed under intersection, (iv) $\Phi$ is closed under converse (i.e., exchanging the two argument places), and (v) for any $V\in\Phi$, there exists a binary relation $W\in\Phi$ for which $W\circ W\subseteq V$, where $W\circ W$ is a composition $\{(v,w)\in V^2 \mid \exists z\in V\,[(v,z),(z,w)\in W]\}$. For more details, refer to, e.g., \cite[Chapter II]{Bou66}.
A simple example of such uniform structures is given by $\Phi_{\real} =\{U_{\varepsilon}\mid \varepsilon >0\}$ on $V=\real$, where $U_{\varepsilon} = \{(x,y)\in V^2\mid |x-y|< \varepsilon\}$ for any $\varepsilon\geq0$. It is not difficult to show that Conditions (i)--(v) hold for $\Phi_{\real}$. To see Condition (v), for instance, for any given $V = U_{\varepsilon}$ in $\Phi_{\real}$, if we take  $W=U_{\varepsilon'}$ in $\Phi_{\real}$ with $\varepsilon'=\frac{\varepsilon}{2}$,
then $W\circ W\subseteq V$ obviously follows.

A uniform structure $\Phi$ on $V$ is said to be \emph{compatible} with a given  topology $T_V$ if, for every set $A\subseteq V$,  $A\in T_V$ holds exactly when, for every $x\in A$, a certain set $U\in\Phi$ satisfies $U[x] \subseteq A$, where $U[x]=\{y\in V\mid (x,y)\in U\}$. A topological space $(V,T_V)$ is \emph{uniformizable} if there exists a uniform structure compatible with the topology $T_V$. For example, the aforementioned $\Phi_{\real}$ is compatible with the standard topology on $\real$ whose basis consists of all neighborhoods of the form $N_{\varepsilon}(x)=\{y\in\real\mid (x,y)\in |x-y|<\varepsilon\}$ for any $x\in\real$ and $\varepsilon\geq0$, because $N_{\varepsilon}(x)$ is expressed as $U_{\varepsilon}[x] =\{y\in\real\mid (x,y)\in U_{\varepsilon}\}$.
With respect to the set $C(V)$ of all continuous maps on $V$, a subset $F$ of $C(V)$ is \emph{uniformly topologically equicontinuous} if, for any element $U$ of a uniform structure $\Phi$ on $V$, the set $\{(u,v)\in V^2\mid \forall f\in F\,[(f(u),f(v))\in U]\}$  belongs to $\Phi$.

Recall from Section \ref{sec:automata-base} that, for an automata base $(\VV,\BB,\OO)$, $\BB$ is composed of all subsets $F$ of $C(V)$ for any  space $V\in\VV$.
We say that $\VV$ is \emph{compact} if every topological space
$(V,T_V)$ in $\VV$ is compact.
We further say that a sub-automata base $(\VV,\BB)$ is \emph{uniformly topologically equicontinuous} if, for any space $V\in\VV$ and any subset $F_V$ of $C(V)$ in $\BB$, $F_V$ is uniformly topologically  equicontinuous.

In what follows, let us present our assertion on a natural condition on $(\VV,\BB,\OO)$ that can ensure $(\VV,\BB,\OO)\mbox{-}\mathrm{1DTA} = \reg$. This gives a complete characterization of regular languages in terms of topological automata.

\begin{ytheorem}\label{iff-cond-uniform}
For any language $L$, the following two statements are logically equivalent.
\begin{enumerate}
  \setlength{\topsep}{-2mm}%
  \setlength{\itemsep}{1mm}%
  \setlength{\parskip}{0cm}%

\item $L$ is regular.

\item There is an automata base $(\VV,\BB,\OO)$ such that every element in $\VV$ is uniformizable, $\VV$ is compact, $(\VV,\BB)$ is uniformly topologically  equicontinuous, $\OO$ is closed under inverse images of maps with respect to $\BB$, and $L$ is recognized by a certain $(\VV,\BB,\OO)$-1dta.
\end{enumerate}
\end{ytheorem}

\begin{proof}
(1 $\Rightarrow$ 2)
Any 1dfa can be viewed as a $(\VV,\BB,\OO)$-1dta of a particular form described in Section \ref{sec:1dta-characterize}. In the description of this 1dta, all elements in $\VV$ are uniformizable and its sub-automata base $(\VV,\BB)$ is compact and uniformly topologically equicontinuous.

(2 $\Rightarrow$ 1)
Take any language $L$ over an alphabet $\Sigma$. Let $(\VV,\BB,\OO)$ be any automata base for which $\OO$ is closed under inverse images of maps with respect to $\BB$.
We assume that $\VV$'s elements $V$ are all uniformizable, $\VV$ is compact, and $(\VV,\BB)$ is uniformly topologically equicontinuous. By the uniformizability of $V$, there exists a uniform structure $\Phi_M$ of $V$ that is compatible with the topology $T_{V}$ on $V$.
Assume that there is a $(\VV,\BB,\OO)$-1dta $M =  (\Sigma,\{\cent,\dollar\},V,\{B_{\sigma}\}_{\sigma\in\check{\Sigma}}, v_0,E_{acc},E_{rej})$ that recognizes $L$.
In what follows, we intend to prove that $L$ is a regular language by converting $M$ into its equivalent 1dfa $N$. Clearly, $V$ is in $\VV$ and there is a subset $G_V$ of $C(V)$ in $\BB$ satisfying that $\{B_{x}\}_{x\in \Sigma_{\cent\dollar}^*} \subseteq  G_V$, where $\Sigma_{\cent\dollar}^*
= \Sigma^*\cup \{\cent\} \Sigma^*\{\dollar,\lambda\}$.

To simplify our proof, we first eliminate the right-endmarker $\dollar$ from $M$.
For this purpose, as in the proof of Lemma \ref{right-endmarker}, we define $E^{\dollar}_{acc} = B^{-1}_{\dollar}(E_{acc})$ and $E^{\dollar}_{rej} = B^{-1}_{\dollar}(E_{rej})$. By the closure property of $\OO$, $(E^{\dollar}_{acc},E^{\dollar}_{rej})$ obviously belongs to $\OO$. We also need to modify $\Phi_{M}$ to the set $\tilde{\Phi}_{M} = \{A\cap (E^{\dollar}_{\tau}\times E^{\dollar}_{\tau}) \mid \tau\in\{acc,rej\},A\in\Phi_{M}\}$.

Next, we partition $\Sigma^*$ into equivalence classes in the following way. Given two strings $x,y\in\Sigma^*$, we write $x\simeq y$ if $L(xz)=L(yz)$ holds for all  strings $z\in\Sigma^*$. Since $\simeq$ is an equivalence relation, we can consider the collection $\Sigma^*\!/\!\!\simeq$ of all equivalence classes. If $|\Sigma^*\!/\!\!\simeq\!\!|=1$, then either $L=\Sigma^*$ or $L=\setempty$ holds, and thus $L$ is obviously regular. In the following argument, we assume that $|\Sigma^*\!/\!\!\simeq\!\!|>1$.
Let us choose two strings $x$ and $y$ from different equivalence classes to ensure $x\not\simeq y$. For simplicity, we write $\bar{v}_z$ for $B_{\cent z}(v_0)$ for any $z\in\Sigma^*$.
We then define a set $C_{x,y} = \{A\in \tilde{\Phi}_{M}\mid \exists z\in\Sigma^*[(B_z(\bar{v}_x),B_z(\bar{v}_y))\notin A]\}$ and consider a ``maximal'' set $D_{x,y}$ in $C_{x,y}$ in the sense that, for any $A\in C_{x,y}$, $D_{x,y}\subseteq A$ implies $D_{x,y} =A$.
From this set $D_{x,y}$, we define the set $F_{x,y}= \{(v_1,v_2)\in V^2 \mid \forall z\in\Sigma^*[(B_z(v_1),B_z(v_2))\in D_{x,y}]\}$. Obviously, we obtain $(v,v)\in F_{x,y}$ for any $v\in V$ but $(\bar{v}_x,\bar{v}_y)\notin F_{x,y}$. Since $G_V$ is uniformly topologically equicontinuous, $F_{x,y}$ falls into $\tilde{\Phi}_{M}$.

Furthermore, we set $P=\{ (u,F_{x,y})\mid x,y\in\Sigma^*,  x\not\simeq y, \exists w[ (u,w)\in F_{x,y}] \}$ and claim that $V =\bigcup_{(u,F)\in P}F[u]$, where $F[u] = \{y\in V\mid (u,y)\in F\}$. From $F[u]\subseteq V$, it follows that $\bigcup_{(u,F)\in P} F[u] \subseteq V$. Thus, it suffices to show that $V\subseteq \bigcup_{(u,F)\in P}F[u]$. Let us choose two strings $x$ and $y$ satisfying $x\not\simeq y$. For any $v\in V$, since $(v,v)\in F_{x,y}$, we obtain $(v,F_{x,y})\in P$ and thus $v\in F_{x,y}[v]$.
As a consequence, $\{F[u]\mid (u,F)\in P\}$ turns out to be a covering of $V$. Therefore, we conclude that $V = \bigcup_{(u,F)\in P} F[u]$.

By the compactness of $V$, from $\{F[u]\mid (u,F)\in P\}$, we choose  a finite subcovering $\{P_i\}_{i\in[t]}$ of $V$, where $t$ is a certain number in  $\nat^{+}$. Let us consider all possible nonempty intersections of an arbitrary number of sets in $\{P_i\}_{i\in[t]}$ and define $\PP$ as the set of all such intersections.
Next, we claim that (*) for any $A\in\PP$ and for any two strings $x,y\in\Sigma^*$ with $\bar{v}_x,\bar{v}_y\in A$, both $x$ and $y$ belong to the same equivalence class; namely, $x\simeq y$. To show this claim, we assume, on the contrary, that
$\bar{v}_x,\bar{v}_y\in A$ but $x\not\simeq y$.
This assumption implies the existence of a string $z$ for which $(B_z(\bar{v}_x),B_z(\bar{v}_y))$ belongs to $(E^{\dollar}_{acc}\times E^{\dollar}_{rej}) \cup (E^{\dollar}_{rej}\times E^{\dollar}_{acc})$.
The definition of $D_{x,y}$ implies that $(B_z(\bar{v}_x),B_z(\bar{v}_y))\notin D_{x,y}$. It then follows that $(\bar{v}_x,\bar{v}_y)\notin F_{x,y}$. Concerning the set $A$, we take a pair $(u,F)\in P$ satisfying $A\subseteq F[u]$.
Since $\bar{v}_x,\bar{v}_y\in A$, we obtain $(\bar{v}_x,u),(u,\bar{v}_y)\in F$.
Since $F\in\tilde{\Phi}_{M}$ and $\Phi_M$ is a uniform structure, there exists a set $X\in\Phi_M$ such that $(\bar{v}_x,u),(u,\bar{v}_y)\in X$ implies $(\bar{v}_x,\bar{v}_y)\in F$. Since $F\subseteq F_{x,y}$, we conclude that $(\bar{v}_x,\bar{v}_y)\in F_{x,y}$.
This is a clear contradiction against $(\bar{v}_x,\bar{v}_y)\notin F_{x,y}$.
Letting $m=|\PP|$, we express $\PP$ as $\{P'_1,P'_2,\ldots,P'_m\}$, where $P'_1$ must contain $v_0$. For each index $i\in[m]$, we choose a point, say, $v_{i-1}$ that represents $P'_i$, and we define $Q=\{v'_0,v'_1,\ldots,v'_{m-1}\}$.
For any pair $u,w\in V$, we write $u\equiv w$ if there exists an index $i\in[m]$ for which $u$ and $w$ are in the same set $P'_i$.

As the final step, the desired 1dfa $N=(Q,\Sigma,\{\cent,\dollar\},\delta,v'_0,Q_{acc},Q_{rej})$ is defined in the following way. Let $Q_{acc}= Q\cap E_{acc}$ and $Q_{rej}= Q\cap E_{rej}$, and define $\delta$ as follows: $\delta(v'_i,\sigma)=v'_j$ iff there exists a point $w_j\in V$ such that $B_{\sigma}(v'_i)=w_j$ and $w_j \equiv v'_j$.
In what follows, we prove that $\delta$ is a well-defined function from $Q \times \check{\Sigma}$ to $Q$. Assume that $\delta(v'_k,\sigma) =v'_i$ and $\delta(v'_k,\sigma)=v'_j$. There are two points $w_i$ and $w_j$ satisfying
that  $B_{\sigma}(v'_k)=w_i$,  $w_i\equiv v'_i$, $B_{\sigma}(v'_k)=w_j$, and $w_j\equiv v'_j$. Since $B_{\sigma}$ is a function, we obtain $w_i=w_j$, which  implies $v'_i\equiv v'_j$. By the definition of $v'_i$ and $v'_j$, we conclude that $v'_i=v'_j$. Therefore, $\delta$ is well-defined.

It follows from the definition of $N$ that $M$ accepts (resp., rejects) $x$ iff $N$ accepts (resp., rejects) $x$. Therefore, $L(M)=L(N)$ follows.
Since $N$ is a 1dfa, $L$ must be a regular language.
\end{proof}


The compactness condition used in Theorem \ref{iff-cond-uniform} is, in fact, an essential assumption for the theorem because, without the compactness, 1dta's may have infinite configuration spaces, which make the 1dta's recognize non-regular languages, as shown in the following lemma.

\begin{ylemma}\label{non-compact-regular}
Let $\VV=\{(\integer,T_{\integer})\}$, $\BB$ consists of a set $F=\{B_x\}_{x\in\Sigma^*_{\cent\dollar}}$ over $\Sigma=\{a,b\}$, and $\OO=\{(E_{acc},E_{rej})\}$, where $T_{\integer} = \PP(\integer)$,  $B_{\cent}=B_{\dollar}=I$, $B_a(n)=n+1$, $B_{b}(n)=n-1$ for all $n\in\integer$, $E_{acc}=\{0\}$, and $E_{rej}=\integer-\{0\}$. The sub-automata base $(\VV,\BB)$ is  uniformly topologically equicontinuous but not compact. There exists a $(\VV,\BB,\OO)$-1dta that recognizes the language $Equal=\{w\in\{a,b\}^*\mid \#_a(w)=\#_b(w)\}$, where $\#_a(w)$ indicates the total number of occurrences of a symbol $a$ in a given string $w$.
\end{ylemma}

\begin{proof}
Take $\VV$, $\BB$, $\OO$, $\Sigma$, and $F$ as in the premise of the lemma. Firstly, we note that $(\integer,T_{\integer})$ is not compact because the set $W=\{\{n\}\mid n\in\integer\}$ is a covering of $\integer$ but no finite subcovering exists for $W$. Obviously, $F$ is a proper subset of $C(\integer)$.
We define $\Phi$ to be the collection of all sets $\{(m,n)\in\integer^2\mid |n-m|\leq k\}$ for any number $k\in\integer$ as well as their \emph{super sets}. It is not difficult to show that (1) $\Phi$ is a uniform structure on $\integer$, (2) $\Phi$ is compatible with $T_{\integer}$, and (3) $F$ is uniformly topologically equicontinuous.
As a consequence, we conclude that $(\VV,\BB)$ is uniformly topologically  equicontinuous.

Let us consider a $(\VV,\BB,\OO)$-1dta $M=(\Sigma,\{\cent,\dollar\},\integer,\{B_{\sigma}\}_{\sigma\in\check{\Sigma}}, v_0,E_{acc},E_{rej})$ with $v_0=0$, where $\Sigma$ and $B_{\sigma}$'s are all given in the premise of the lemma. Take any input string $x=x_1x_2\cdots x_n$ of length $n$ over $\Sigma$. By the definition of $B_{\sigma}$'s, it is not difficult to show that $B_{\cent x\dollar}(v_0) = \#_{a}(x)-\#_{b}(x)$. Hence, it follows that $B_{\cent x\dollar}(v_0)\in E_{acc}$ iff $\#_a(x)=\#_b(x)$. We thus deduce that $M$ recognizes $Equal$.
\end{proof}

\section{Computational Strengths of Properties on Topological Spaces}\label{sec:compare-properties}

The behaviors of topological automata reflect chosen topological spaces and continuous maps. Since those topological concepts are described by  ``properties'' (or ``features'') of topologies. An example of such properties is the \emph{Hausdorff separation axiom}. It is thus possible to compare the strengths of two different properties of topological spaces by evaluating the computational power of the corresponding topological automata.
For our purpose, it is ideal to disregard the two endmarkers for a general treatment of such properties because the endmarkers are quite different in behavior from other standard input symbols. Therefore, unlike the other sections, we intend to use ``markless 1dta's'' (which have no endmarker, discussed in Section \ref{sec:elimination}) throughout this section.

\subsection{Slim Topological Automata}\label{sec:normalization}

To design finite automata, it is sometimes imperative to make them ``small'' enough. Such a requirement often gives rise to a notion of ``minimal'' finite automata.
For instance, Ehrig and K\"{u}hnel \cite{EK74} earlier discussed the minimality of their topological automata founded on compactly generated Hausdorff metric spaces, where a \emph{compactly generated space} is a topological space $V$ such that every subset $A$ of $V$ is open iff $A\cap C$ is open for any compact subspace $C\subseteq V$.
From a different viewpoint, Jeandel \cite{Jea07} considered ``small'' topological automata  under the term of ``purge'' by excluding all points of a given configuration space that cannot be reached (or visited) along any computation.
We wish to take a similar approach to leave out all unreachable points from every topological space.

To be more concrete, consider a markless  $(\VV,\BB,\OO)$-1dta  $M=(\Sigma,V,\{B_{\sigma}\}_{\sigma\in{\Sigma}}, v_0,E_{acc},E_{rej})$ for a given automata base $(\VV,\BB,\OO)$.
Let $F$ denote an appropriate subset of $C(V)$ in $\BB$ containing all maps $B_x$ for any $x\in\Sigma^*$.
There may be a case where all configurations generated (or visited) by $M$ starting with $v_0$ do not cover all points in $V$. In such a case, the topological feature of $V$ does not seem to represent the actual behavior of $M$, because all the points that are unreachable by $M$ may possibly satisfy a completely different property from the rest of the points.
Therefore, to discuss the true power of topologies used to define 1dta's, it is   desirable to leave out all the points that are unreachable by $M$ and to stay focused  on the set of all the points that $M$ can visit.

As a quick example, let us consider two topological spaces $(V_1,T_{V_1})$ and $(V_2,T_{V_2})$, where $V_1=\{v_0,v_1\}$, $T_{V_1} = \{\setempty,\{v_0\},\{v_1\},V_1\}$, $V_2= \{v_0,v_1,v_2\}$, and $T_{V_2} = \{\setempty,\{v_0\},\{v_1\},\{v_0,v_1\},V_2\}$. Obviously, $T_{V_1}$ is the discrete topology but $T_{V_2}$ is not. Let $\Sigma=\{a\}$ and choose a map  $B_{a}$ defined as $B_a(v_0)=v_1$ and $B_a(v_1)=v_0$.
Although $V_1$ and $V_2$ are quite different, any markless  1dta having $B_{a}$ behaves in the same way on $V_1$ and $V_2$ since $v_2$ is not reachable  from $v_0$.

The above argument makes us introduce a new notion of  ``slim 1dta's,'' which have no endmarker and visit all points in $V$.
Formally, a \emph{slim 1dta} is a markless 1dta such that, for every point $v\in V$, there exists a string $x\in\Sigma^*$ satisfying $B_{x}(v_0)=v$.
In what follows, we wish to present how to construct, from any given markless 1dta $M$,  its equivalent slim 1dta.
The \emph{normalization} of $M$ is defined to be a markless $(\hat{\VV}_M,\hat{\BB}_M,\hat{\OO}_M)$-1dta, denoted by $M_{norm}$, which is obtained by modifying $M$ in the following way.
Firstly, we define $F'$ to be the set $\{B_{x}\}_{x\in\Sigma^*}$ and further define $V_{M}= \{v_0, B(v_0) \mid B\in F' \}$ with a subspace topology on $V_M$ induced from $T_{V}$.
Notice that $(V_M,T_{V_M})$ and $(V,T_V)$ may be quite different in nature.
We further set $\hat{\VV}_{M}=\{V_{M}\}$.
To define $\hat{\BB}_M$, we need to restrict the domain of each map $B$ in $F'$ onto $V_M$.
Recall from Section \ref{sec:basic-model} that such a restricted map is expressed as $B|V_M$. We define $F_M$ to be the set $\{B|V_M\mid B\in F'\}$. The desired $\hat{\BB}_M$ is then set to be $\{F_M\}$.
Finally, we set $\hat{\OO}_M$ to be
$\{(E_{acc}\cap V_{M},E_{rej}\cap V_{M}) \}$. For each script $\tau\in\{acc,rej\}$, since $E_{\tau}$ is a clopen set, $E_{\tau}\cap V_{M}$ is also clopen with respect to $T_{V_M}$.

\begin{ylemma}\label{slim-equal}
For any markless $(\VV,\BB,\OO)$-1dta $M$, let $M_{norm}$ be the normalization of $M$. The following properties hold for $M_{norm}$.
\begin{enumerate}
  \setlength{\topsep}{-2mm}%
  \setlength{\itemsep}{1mm}%
  \setlength{\parskip}{0cm}%

\item $M_{norm}$ is slim.

\item $M_{norm}$ is computationally equivalent to $M$.
\end{enumerate}
\end{ylemma}

\begin{proof}
(1) To show the slimness of $M_{norm}$, let $v$ be any configuration in $V_M$. Consider the set $\BB'$ and its element $F'$.
There exists a map $B\in F'$ for which  $v=  B(v_0)$.
By the definition of $F_M$, we can take a string $x\in\Sigma^*$ for which $B=B_x|V_M$.
Since $v_0,v\in V_M$, we obtain $v=B_{x}|V_M(v_0)$.
It is thus clear that all points of $V_{M}$ are visited by $M$ while reading certain input strings over the alphabet $\Sigma$; therefore, $M_{norm}$ is slim.

(2) We want to show by induction on $n\in\nat$ that,  for any string $x\in\Sigma^n$, $\hat{B}_{x}(v_0)= B_{x}(v_0)$ holds, because
this result establishes the computational equivalence between $M_{norm}$ and $M$.
Take any string of the form $x\sigma\in\Sigma^{n+1}$ for a symbol $\sigma\in\Sigma$ and consider $B_{x\sigma}(v_0)$. Note that $v_0\in V_M$ by the definition of $V_M$.  Assume by induction hypothesis that $\hat{B}_{x}(v_0) = B_{x}(v_0)\in V_M$.
Since $\hat{B}_{\sigma}$ is a restriction of $B_{\sigma}$ onto $V_M$, $\hat{B}_{\sigma}(w) = B_{\sigma}(w)$ holds for any $w\in V_M$. It then follows that $\hat{B}_{x\sigma}(v_0) = \hat{B}_{\sigma}(\hat{B}_{x}(v_0)) = \hat{B}_{\sigma}(B_{x}(v_0)) = B_{\sigma}(B_{x}(v_0)) = B_{x\sigma}(v_0)$.
\end{proof}

\begin{ylemma}
Given an automata base $(\VV,\BB,\OO)$, if a markless  $(\VV,\BB,\OO)$-1dta $M = (\Sigma,V, \{B_{\sigma}\}_{\sigma\in\check{\Sigma}}, v_0,E_{acc},E_{rej})$ is slim and $\BB$ contains a superset of the set $F = \{B_x\}_{x\in\Sigma^*}$  as its element, then $M_{norm}$ is also a markless  $(\VV,\BB,\OO)$-1dta.
\end{ylemma}

\begin{proof}
Let $(\VV,\BB,\OO)$ and $M$ be given as in the premise of the lemma. Let us recall  that $M_{norm}$ is a markless $(\hat{\VV}_{M},\hat{\BB}_{M},\hat{\OO}_{M})$-1dta induced from $(\VV,\BB,\OO)$ and $M$ with $\hat{\VV}_M=\{V_M\}$ and $\hat{\BB}_M=\{F_M\}$.
It thus suffices to show that $\hat{\VV}_{M}\subseteq \VV$, $\hat{\BB}_{M}\subseteq \BB$, and $\hat{\OO}_{M}\subseteq \OO$.
Since $M$ is slim, we obtain $V_M=V$ together with $T_{V_M}=T_V$. Thus, $\hat{\VV}_{M}\subseteq \VV$ follows.
Recall the set $F'$, which induces $F_M$ and $V_{M}$.
For any map  $\hat{B}$ in $F_M$, there is another map $B'\in F'$ for which  $\hat{B}$ equals $B'$ restricted to $V_M$, namely, $\hat{B}=B'|V_M$. Since $V_M=V$, we obtain $\hat{B}=B'$. This yields the desired inclusion $\hat{\BB}_{M}\subseteq \BB$.  Moreover, it follows that $\hat{\OO}_M = \{(E_{acc}\cap V_M,E_{rej}\cap V_M)\} = \{(E_{acc},E_{rej})\} \subseteq \OO$. Therefore, $M_{norm}$ is a markless  $(\VV,\BB,\OO)$-1dta.
\end{proof}

\subsection{Computational Strengths of Topological Features}

With the use of slim 1dta's, we intend to compare the strengths of topological features by evaluating the computational power of the associated slim 1dta's. Given a property $P$ of topologies in question, we say that an automata base $(\VV,\BB,\OO)$ \emph{meets} $P$ if  every slim $(\VV,\BB,\OO)$-1dta $M$ satisfies $P$.
Let $P_1$ and $P_2$ be two properties of topologies. We say that $P_2$ \emph{supersedes} $P_1$, denoted by $P_1\sqsubseteq P_2$, exactly when every automata base $(\VV,\BB,\OO)$ that meets $P_1$ also meets $P_2$.
Furthermore, we say that $P_2$ is \emph{at least as computationally strong as} $P_1$, denoted by $P_1\leq_{\mathrm{comp}} P_2$, if, for any automata base $(\VV_1,\BB_1,\OO_1)$ that meets $P_1$, there exists another automata base  $(\VV_2,\BB_2,\OO_2)$ meeting $P_2$ such that every slim $(\VV_1,\BB_1,\OO_1)$-1dta has a computationally equivalent slim $(\VV_2,\BB_2,\OO_2)$-1dta.
Notice that $P_1$ is always at least as computationally strong as itself. Moreover, $P_2$ is said to be \emph{computationally stronger than} $P_1$ if $P_1\leq_{\mathrm{comp}} P_2$ and $P_2\not\leq_{\mathrm{comp}} P_1$. In this case, we succinctly write $P_1<_{\mathrm{comp}} P_2$.

The following lemma is immediate.

\begin{ylemma}
For two properties $P_1$ and $P_2$ of topologies, if $P_1\sqsubseteq P_2$, then $P_1\leq_{\mathrm{comp}}P_2$.
\end{ylemma}

\begin{proof}
Given two properties $P_1$ and $P_2$ of topologies, assume that $P_1\sqsubseteq P_2$. Let us consider any automata base $(\VV,\BB,\OO)$ that  meets $P_1$. Since $P_1\sqsubseteq P_2$, $(\VV,\BB,\OO)$ also meets $P_2$. By the definition of $\leq_{\mathrm{comp}}$, $P_1\leq_{\mathrm{comp}}P_2$ follows immediately.
\end{proof}

Next, we present two results concerning topological indistinguishability, which has been introduced in Section \ref{sec:finite-topology}. Let $(V,T_V)$ be any topological space.
The \emph{Kolmogorov separation axiom} dictates the property that any pair of distinct points of $V$ are topologically distinguishable. Any space that satisfies the Kolmogorov separation axiom is called a \emph{Kolmogorov space}. The discrete topology always satisfies the Kolmogorov separation axiom.
By contrast, the trivial topology is a simple example of topologies that  violate the Kolmogorov separation axiom.
As another example of topological space $(V,T_V)$ with $V=\{1,2,3\}$ and  $T_V=\{\setempty,\{1,2\}, \{2\}, \{2,3\}, \{1,2,3\}\}$,
the topological space $(V,T_V)$ is clearly a Kolmogorov space although  $T_V$  is not the discrete topology.

\begin{yproposition}
The discrete topology is computationally stronger than any topology violating  the Kolmogorov separation axiom.
\end{yproposition}

\begin{proof}
Let us consider any language $L$ over the binary alphabet $\Sigma=\{0,1\}$ satisfying the following condition: for any two distinct strings $x_1$ and $x_2$ over $\Sigma$, there exists a string $y\in\Sigma^*$ for which $L(x_1y) \neq L(x_2y)$. For such a language $L$, we want to prove by contradiction that any slim $(\VV,\BB,\OO)$-1dta violating the Kolmogorov separation axiom cannot recognize $L$. To lead to a contradiction, we assume that $L$ is recognized by a certain slim $(\VV,\BB,\OO)$-1dta $M$ whose topological space $(V,T_V)$ violates  the Kolmogorov separation axiom; that is, there exists a pair of distinct points $v_1,v_2\in V$ that are topologically indistinguishable. Hereafter, we fix these points $v_1$ and $v_2$.
By the slimness of $M$,  the strings $x_1$ and $x_2$ satisfy both  $B_{x_1}(v_0)=v_1$ and $B_{x_2}(v_0)=v_2$.
Since no open set topologically distinguishes between $v_1$ and $v_2$, for any string $y\in\Sigma^*$, $B_{y}(v_1)$ and $B_{y}(v_2)$ cannot be topologically distinguishable and they together fall into the same set, either $E_{acc}$ or $E_{rej}$. Therefore, we conclude that $L(x_1y) = L(x_2y)$ for every string $y$, a contradiction.

Since $L$ is recognized by a certain 1dta with the discrete topology, as shown in Proposition \ref{all-languages}, the proposition follows immediately.
\end{proof}

The next theorem signifies a clear difference in computational strength between the trivial topology and any topology that violates the Kolmogorov separation axiom.

\begin{ytheorem}\label{Kolmogorov}
There is a topology, which is computationally stronger than the trivial topology but does not satisfy the Kolmogorov separation axiom.
\end{ytheorem}

\begin{proof}
Let us consider the language $ZERO=\{0^n\mid n\in\nat\}$ over the binary alphabet $\Sigma=\{0,1\}$. By Proposition \ref{trivial-topology},
for any automata base $(\VV,\BB,\OO)$ whose topological spaces $V$ in $\VV$ have the trivial topology, $ZERO$ is recognized by no $(\VV,\BB,\OO)$-1dta working over the binary alphabet.
From this, we set our goal to construct an automata base $(\VV,\BB,\OO)$ and its slim $(\VV,\BB,\OO)$-1dta $M$ satisfying that (i) $\VV$ consists of finite topological spaces $(V,T_V)$ violating the Kolmogorov separation axiom and (ii) $M$ recognizes $ZERO$.
These conditions make us conclude that the topology on $V$ is computationally stronger than the trivial topology.

Firstly, let us define the desired slim 1dta $M =(\Sigma,V, \{B_{\sigma}\}_{\sigma\in \Sigma}, v_0,E_{acc},E_{rej})$ as follows. Let $V=\{0,1,2\}$ and  $T_V=\{\setempty,V,\{0\},\{1,2\}\}$. Clearly, $(V,T_V)$ violates the Kolmogorov separation axiom.
We then define $B_0=I$ and $B_1(n)=\min\{n+1,2\}$ for any element $n\in V$. Note that, for each symbol $\sigma\in\check{\Sigma}$, $B_{\sigma}$ is continuous. Moreover, we set $v_0=0$, $E_{acc}=\{0\}$, and $E_{rej}=\{1,2\}$. It is not difficult to show that $M$ accepts all strings of the form $0^n$ for any $n\in\nat$ and rejects all the strings containing the symbol $1$. Therefore, $M$ recognizes $ZERO$.

Secondly, we define the desired automata base $(\VV,\BB,\OO)$ as follows. Let $\VV=\{V\}$, let $\BB$ be composed of the closure of the set $\{B_x\}_{x\in\Sigma^*}$ under functional composition,
and let $\OO=\{(E_{acc},E_{rej})\}$. Clearly, $M$ is a markless $(\VV,\BB,\OO)$-1dta. This completes the proof of the theorem.
\end{proof}

\section{Multi-Valued Operators and Nondeterminism}\label{sec:nondeterminism}

\emph{Nondeterminism} is a ubiquitous feature, which appears in many fields of  computer science. Jeandel \cite{Jea07} considered such a feature for his model of topological automata to analyze the behaviors of nondeterministic quantum finite automata. In a similar vein, we wish to define a nondeterministic version of our $(\VV,\BB,\OO)$-1dta's, called \emph{one-way nondeterministic topological automata}
(or 1nta's, for short), in such a way that it naturally extends the standard definition of \emph{one-way nondeterministic finite automata} (or 1nfa's),
each of which nondeterministically chooses  at every step one inner state out of a predetermined set of possible next inner states until certain halting states are reached.

\subsection{Multi-Valued Operators and 1nta's}

Unlike the previous sections, we deal with multi-valued operators, which map one  element to ``multiple'' elements. To be more precise, a \emph{multi-valued operator} is a map from each point $x$ of a given topological space $(V_1,T_{V_1})$ to a number (including ``zero'') of points of another topological space $(V_2,T_{V_2})$. Although this operator can be viewed simply as an ``ordinary'' map from $V_1$ to $\PP(V_2)$, we customarily express such a multi-valued operator as $B:V_1\to V_2$ as long as the multi-valuedness of $B$ is clear from the context.
In comparison, any standard map is referred to as a \emph{single-valued operator}.  Notice that, by the definition, every single-valued operator can be viewed as a multi-valued operator.

As a quick example, let us consider $V=\real$ and the discrete topology $T_V$ on $\real$. Given a constant $\varepsilon>0$, the function $F_{\varepsilon}$ defined by $f_{\varepsilon}(v) = \{w\in V \mid |w-v|\leq\varepsilon\}$ is a multi-valued operator on $V$. Another example is the {inverse operator} defined  in
Section \ref{sec:elimination}.
For any single-valued map $B$ on $V$, if we define
$B^{-1}(v) =\{w\in V\mid B(w)=v\}$ for each point $v\in V$, then this new operator $B^{-1}$ is clearly a multi-valued operator.

Let $B$ denote a multi-valued operator on $V$, namely, $B:V\to V$.
For any subset $A$ of $V$, the notation $B(A)$ denotes the union $\bigcup_{v\in A}B(v)$.
A \emph{neighborhood} of a set $C$ of points in $V$ is the union $\bigcup_{x\in C} N_x$, where each $N_x$ is a neighborhood of a point $x$ in $C$ defined in Section \ref{sec:topology-notion}.
A multi-valued operator $B:V_1\to V_2$ is said to be \emph{continuous} if,  for any $x\in V_1$ and for any neighborhood $N$ of $B(x)$ ($\subseteq V_2$), there exists a neighborhood $N'$ of $x$ satisfying $B(N')\subseteq N$, where $B(N') = \bigcup_{z\in N'} B(z)$.
Given a topological space $V$, $C_m(V)$ denotes the set of all continuous multi-valued operators on $V$.
In Section \ref{sec:topology-notion}, we have used the notation $\circ$ to describe the functional composition between two single-valued continuous maps.
To emphasize the multi-valuedness of operators $B_1$ and $B_2$, in contrast, we express their ``functional composition'' as $B_1\diamond B_2$, which satisfies  $(B_1\diamond B_2)(v) = B_1(B_2(v))$ ($=\bigcup_{z\in B_2(v)}B_1(z)$) for any $v\in V$.

Let us define an \emph{extended automata base} by expanding the notion of automata bases in the following way.

\s
{\bf Extended Automata Base.}
An \emph{extended automata base} is, similar to an automata base, a tuple  $(\VV,\BB,\OO)$ in which $\VV$ is a set of topological spaces, $\OO$ is a set of observable pairs, and $\BB$ is composed of subsets $F$ of $C_m(V)$ for each topological space $(V,T_V)\in\VV$ such that $F$ is closed under functional composition $\diamond$.

\s
Since single-valued operators can be viewed as multi-valued ones, any automata base can be treated as a special case of extended automata bases.


\s
{\bf Definition of 1nta's.}
Given an extended automata base $(\VV,\BB,\OO)$, a \emph{one-way nondeterministic $(\VV,\BB,\OO)$-topological automaton} (or a $(\VV,\BB,\OO)$-1nta, for short)
$M$ is a septuple $(\Sigma,\{\cent,\dollar\},V, \{B_{\sigma}\}_{\sigma\in\check{\Sigma}},v_0,E_{acc},E_{rej})$ similar to a 1dta except that, for each symbol $\sigma\in\check{\Sigma}$, $B_{\sigma}$ is a multi-valued continuous operator on $V$.
This machine $M$ works as follows.
On input $x=x_1x_2\cdots x_n$ (which is given in the form $\cent x\dollar$ on  an input tape), we apply $B^{\diamond}_{\cent x\dollar}$ to $v_0$, where  $B^{\diamond}_{\cent x\dollar} = B_{\dollar}\diamond B_{x_n}\diamond \cdots \diamond B_{x_2} \diamond B_{x_1} \diamond B_{\cent}$.
We say that $M$ \emph{accepts} $x$ if $B^{\diamond}_{\cent x\dollar}(v_0) \cap E_{acc}\neq\setempty$ and that $M$ \emph{rejects}\footnote{It is also possible to relax this requirement of $B^{\diamond}_{\cent x\dollar}(v_0) \subseteq E_{rej}$ to $B^{\diamond}_{\cent x\dollar}(v_0)\cap E_{acc} = \setempty$.} $x$ if $B^{\diamond}_{\cent x\dollar}(v_0) \subseteq E_{rej}$.
The notation $(\VV,\BB,\OO)\mbox{-}\mathrm{1NTA}$ is used to denote the family of all languages recognized by appropriate $(\VV,\BB,\OO)$-1nta's.


Let us demonstrate that the following types of nondeterministic finite automata can be characterized  by certain 1nta's in a natural way.

\s
{\bf (i) Nondeterministic Finite Automata.}
A \emph{one-way nondeterministic finite automaton} (or a 1nfa) is described as  $(\Sigma,\{\cent,\dollar\},V, \{B_{\sigma}\}_{\sigma\in\check{\Sigma}}, v_0,E_{acc},E_{rej})$, where $V$ is of the form $[k]$ for a certain constant $k\in\nat^{+}$, $B_{\sigma}(v)$ is a subset of $V$ for each $\sigma\in \check{\Sigma}$, and $E_{acc}$ and $E_{rej}$ are  disjoint nonempty subsets of $V$.

\s
{\bf (ii) Nondeterministic Pushdown Automata.}
A \emph{one-way nondeterministic pushdown automaton} (or a 1npda) is expressed as $(\Sigma,\{\cent,\dollar\},V, \{B_{\sigma}\}_{\sigma\in\check{\Sigma}}, v_0,E_{acc},E_{rej})$, where $V=[k]\times \bot\Gamma^*$ for a constant $k\in\nat^{+}$ and a fixed alphabet $\Gamma$. We demand that $B_{\sigma}(q,\bot z)\subseteq [k]\times \{\bot z_1z_2\cdots z_{n-1}w\mid w\in\Gamma^{\leq l}\}$, where $z=z_1z_2\cdots z_n\in\Gamma^n$ and $l\in\nat^{+}$. Finally, we set  $E_{acc} = Q_1\times \bot\Gamma^*$ and $E_{rej}= Q_2\times \bot\Gamma^*$ with a partition $(Q_1,Q_2)$ of $[k]$.

\s
{\bf (iii) Quantum Interactive Proof Systems with Quantum Finite Automata \cite{NY09,NY15}.}
A \emph{quantum interactive proof (QIP) system with a 1qfa verifier} is, roughly, a 2-player communication game between an adversarial almighty prover and a 1qfa verifier, who interact with each other through a shared common message board holding a single letter. For a positive instance, the honest prover must provide a ``valid'' proof (i.e., a valid piece of information) and the verifier confirms its correctness with high confidence. On the contrary, for a negative instance, no matter which proof a cheating prover provides, the verifier refutes it with high confidence. For ease of description, we assume that a prover behaves classically.
Such a QIP system can be described as a $(\VV,\BB,\OO)$-1nta that satisfies the following conditions. Let $\VV$ contain $V_1\times V_2$, where $V_1$ consists of $k_1$-dimensional normalized basic vectors and $V_2$ is $(\complex^{\leq 1})^{k_2}$ for certain constants $k_1,k_2\in\nat^{+}$.
The set $\BB$ consists of sets $F$, each of which is a collection of multi-valued operators $B_{\sigma}: V_1\times V_2\to V_1\times V_2$ for each $V_1\times V_2\in \VV$ such that there are a multi-valued operator $B_{1,\sigma}:V_1\to V_1$ and a single-valued operator $B_{2,\sigma}:V_1\times V_2\to V_1\times V_2$ satisfying $B_{\sigma}(a,v) = B_{2,\sigma}(B_{1,\sigma}(a),v)$ for any $(a,v)\in V_1\times V_2$. Moreover, $\OO$ contains all pairs $(E_{acc},E_{rej})$, where $E_{acc} = \{(a,v)\in V_1\times V_2\mid \|\Pi_{acc}v\|_2^2\geq 1-\varepsilon\}$ and $E_{rej} = \{(a,v)\in V_1\times V_2\mid \|\Pi_{rej}v\|_2^2\geq 1-\varepsilon\}$ for a certain $\varepsilon\in[0,1/2)$.


\ms
Given a multi-valued operator $B$ on $V$, we define its \emph{(multi-valued) inverse operator} $B^{-1}$ as $B^{-1}(v) = \{w\in V\mid v\in B(w)\}$ for every point $v\in V$. We further extend $B^{-1}$ to any subset $A$ of $V$ by setting $B^{-1}(A) = \bigcup_{v\in A}B^{-1}(v)$.
We say that a set $F$ of multi-valued operators is \emph{closed under inverse} if, for any $B\in F$, the (multi-valued) inverse operator $B^{-1}$ belongs to $F$. Furthermore, a set $\BB$ of families of multi-valued operators is said to be \emph{closed under inverse} if every set $F$ in $\BB$ is closed under inverse.

The following lemma provides basic features of (multi-valued) inverse operators.

\begin{ylemma}\label{inverse-basic}
Given a topological space $(V,T_V)$ and a multi-valued operator $B$ on $V$ (i.e., $B:V\to V$), it  follows that, for any nonempty sets $A,A_1,A_2\subseteq V$, (1) $A \subseteq  (B^{-1}\diamond B)(A)$ and $A\subseteq (B\diamond B^{-1})(A)$, (2) if $A_1\cap B^{-1}(A_2) \neq\setempty$, then $B(A_1)\cap A_2 \neq\setempty$,
and (3) if $A_1\subseteq A_2$, then $B(A_1)\subseteq B(A_2)$ and $B^{-1}(A_1)\subseteq B^{-1}(A_2)$.
\end{ylemma}

\begin{proof}
(1) We begin with the first claim. Given a point $v\in A$, let $C_v =B(v)$. It then follows that $B^{-1}(C_v) = \bigcup_{z\in C_v}B^{-1}(z) = \{ w\in V\mid \exists z\in C_v[z\in B(w)]\}$, which equals $\{w\in V\mid \exists z\in V [z\in B(v)\cap B(w)]\}$. The last expression clearly indicates that $v\in B^{-1}(C_v)$. Hence, we conclude that $A\subseteq (B^{-1}\diamond B)(A)$.

For the second claim, let $v\in A$ and set $D_v=B^{-1}(v)$. Note that $B(D_v) = \bigcup_{z\in D_v} B(z) = \{w\in V\mid \exists z\in D_v[w\in B(z)]\}$. Hence, $B(D_v)$ equals $\{w\in V\mid \exists z\in V[v\in B(z)\wedge w\in B(z)]\}$. From this expression, we conclude that $B(D_v)$ contains $v$. Consequently, we obtain $A\subseteq (B\diamond B^{-1})(A)$.

(2) For any point $v\in A_1\cap B(A_2)$, we take another point $z\in A_2$ for which $z\in B(v)$. Since $z\in B(A_1)$, it follows that $z\in B(A_1)\cap A_2$. Therefore, we obtain $B(A_1)\cap A_2\neq\setempty$.

(3) This is trivial from the definition of $B$ and $B^{-1}$.
\end{proof}

Hereafter, we present a simple observation on the closure property under reversal.
A language family $\CC$ is said to be \emph{closed under reversal} if, for any language $L\in\CC$, its reversal $L^{R}$ ($=\{x\mid x^R\in L\}$)
also belongs to $\CC$.
Although $\reg$ is known to be closed under reversal, 1dta's in general do not support this closure property.

We begin with a quick preparation for our observation (Proposition  \ref{reversal-closure}).
Given a 1nta with $V$ and $(E_{acc},E_{rej})$, we choose two points $v_{acc}\in E_{acc}$ and $v_{rej}\in E_{rej}$, and we then define a \emph{single-valued} operator $D_V[v_{acc},v_{rej}]:V\to V$ as $D_V[v_{acc},v_{rej}](v)=v_{acc}$ if $v\in E_{acc}$; $v_{rej}$ if $v\in E_{rej}$; and $v$ otherwise.
The use of $D_V[v_{acc},v_{rej}]$ helps us fix unique accepting and rejecting  configurations no matter which inputs are given. Note that $D_V[v_{acc},v_{rej}]$ is a continuous operator because of $E_{acc}\cap E_{rej}=\setempty$.

\begin{yproposition}\label{reversal-closure}
Let $(\VV,\BB,\OO)$ be any extended automata base such that $\BB$ is closed under inverse. Assume that, for any $V\in\VV$, a certain subset $F$ of $C_m(V)$ in $\BB$ contains  all operators of the form  $D_V[v_{acc},v_{rej}]$ and $D_V[v_{acc},v_{rej}]\diamond B$ for any multi-valued operator $B\in F$ and for any pair $(v_{acc},v_{rej})\in E_{acc}\times E_{rej}$. If there is a $(\VV,\BB,\OO)$-1nta $M$ with $v_0$, $V$, and $(E_{acc},E_{rej})$ satisfying $\{v_0\}, V-\{v_0\}\in T_{V}$, then there exists a $(\VV,\BB,\OO)$-1nta $N$ that recognizes the reversal of $L(M)$.
\end{yproposition}

\begin{proof}
Take an extended automata base $(\VV,\BB,\OO)$ satisfying the premise of the lemma. Let $M = (\Sigma,\{\cent,\dollar\}, V,\{B_{\sigma}\}_{\sigma\in\check{\Sigma}}, v_0,E_{acc},E_{rej})$ be any $(\VV,\BB,\OO)$-1nta and define $L=L(M)$.
For our later argument, we fix a halting-configuration pair  $(v_{acc},v_{rej}) \in E_{acc} \times E_{rej}$.
We then set $v'_0=v_{acc}$, $\tilde{E}_{acc}=\{v_0\}$, and $\tilde{E}_{rej}= V-\{v_0\}$. Since $\{v_0\},V-\{v_0\}\in T_V$, both $\tilde{E}_{acc}$ and $\tilde{E}_{rej}$ are clopen sets. Moreover, we define
$\tilde{B}_{\cent} = (D_V[v_{acc},v_{rej}]\diamond B_{\dollar})^{-1}$,  $\tilde{B}_{\dollar} = B^{-1}_{\cent}$, and  $\tilde{B}_{\sigma} =B^{-1}_{\sigma}$ for every symbol $\sigma\in\Sigma$.
We then obtain a 1nta $N = (\Sigma,\{\cent,\dollar\}, V,\{\tilde{B}_{\sigma}\}_{\sigma\in\check{\Sigma}}, v'_0,\tilde{E}_{acc},\tilde{E}_{rej})$.

Hereafter, our goal is to verify that $N$ precisely recognizes $L^R$. Toward this goal,
we first claim that, for any length $n\in\nat$, any string $z=z_1z_2\cdots z_n\in\Sigma^n$, and any index $k\in[0,n]_{\integer}$,  (1) if $B^{\diamond}_{\cent z\dollar}(v_0) \cap E_{acc}\neq\setempty$, then $B^{\diamond}_{\cent z_1z_2\cdots z_{k}}(v_0) \subseteq \tilde{B}^{\diamond}_{\cent z_nz_{n-1}\cdots z_{k+1}}(v_{acc})$  and
(2) if $v_0\in \tilde{B}^{\diamond}_{\cent z\dollar}(v_{acc})$, then $B^{\diamond}_{\cent z_nz_{n-1}\cdots z_{k+1}}(v_0) \cap  \tilde{B}^{\diamond}_{\cent z_1z_2\cdots z_k}(v_{acc}) \neq\setempty$, provided that $z_0$ and $z_{n+1}$ are both treated as $\lambda$.
Assuming that the above statements (1)--(2) are true, let us demonstrate
that $L^R = L(N)$.
Let $n\in\nat$ and $x=x_1x_2\cdots x_n\in\Sigma^n$. If $x\in L^R$, then $x^R\in L$, and thus $B^{\diamond}_{\cent x^R\dollar}(v_0)\cap E_{acc}\neq\setempty$. From this, we deduce from Statement (1) with $k=0$ and $z=x^R$ that $B_{\cent}(v_{0}) \subseteq \tilde{B}^{\diamond}_{\cent x^R}(v_{acc})$.
We then apply $B_{\cent}^{-1}$ to both $B_{\cent}(v_{0})$ and $\tilde{B}^{\diamond}_{\cent x^R}(v_{acc})$. Lemma \ref{inverse-basic}(3) implies that $(B^{-1}_{\cent} \diamond B_{\cent}) (v_0) \subseteq B_{\cent}^{-1} (\tilde{B}^{\diamond}_{\cent x^R}(v_{acc})) = \tilde{B}^{\diamond}_{\cent x\dollar}(v'_0)$ since $\tilde{B}_{\dollar} = B^{-1}_{\cent}$. By Lemma \ref{inverse-basic}(1), it follows that $\{v_0\}\subseteq (B^{-1}_{\cent}\diamond B_{\cent})(v_0)$.
This concludes that $\tilde{B}^{\diamond}_{\cent x\dollar}(v'_0)\cap \tilde{E}_{acc}\neq\setempty$. Therefore, $N$ accepts $x$.

On the contrary, when $x\notin L^R$, since $x^R\notin L$, we obtain $B^{\diamond}_{\cent x^R\dollar}(v_0)\subseteq E_{rej}$. We wish to show that $N$ rejects $x$; in other words, $\tilde{B}^{\diamond}_{\cent x\dollar}(v'_0)\subseteq \tilde{E}_{rej}$. Toward a contradiction, we assume that $\tilde{B}^{\diamond}_{\cent x\dollar}(v'_0)\nsubseteq \tilde{E}_{rej}$.
This implies that $v_0\in \tilde{B}^{\diamond}_{\cent x\dollar}(v_{acc})$.
By setting $k=0$ and $z=x$, Statement (2) leads to the conclusion that $B^{\diamond}_{\cent x^R}(v_0)\cap \tilde{B}_{\cent}(v_{acc}) \neq\setempty$. If we take $B= D_V[v_{acc},v_{rej}]\diamond B_{\dollar}$ in Lemma \ref{inverse-basic}(2), then we deduce that $D_V[v_{acc},v_{rej}](B^{\diamond}_{\cent x^R\dollar}(v_0)) \cap \{v_{acc}\} \neq\setempty$.
This is equivalent to $v_{acc}\in B^{\diamond}_{\cent x^R\dollar}(v_0)$. This contradicts $B^{\diamond}_{\cent x^R\dollar}(v_0) \subseteq E_{rej}$.  Therefore, $N$ must reject $x$.

To complete the proof of the proposition, we still need to verify Statements (1)--(2). We begin with proving Statement (1) by downward induction, provided that  $B^{\diamond}_{\cent z\dollar}(v_0) \cap E_{acc}\neq\setempty$ holds.
In the base case of $k=n$, since $B^{\diamond}_{\cent z\dollar}(v_0)\cap E_{acc}\neq \setempty$, it follows that  $D_V[v_{acc},v_{rej}] (B^{\diamond}_{\cent z\dollar}(v_0)) = \{v_{acc}\}$. We then apply $\tilde{B}_{\cent}$ ($= (D_V[v_{acc},v_{rej}]\diamond B_{\dollar})^{-1}$)   and obtain $B^{\diamond}_{\cent z}(v_0)\subseteq \tilde{B}_{\cent}(v_{acc})$ by Lemma \ref{inverse-basic}(1)\&(3).
By induction hypothesis, we obtain $B^{\diamond}_{\cent z_1\cdots z_{k+1}}(v_0) \subseteq \tilde{B}^{\diamond}_{\cent z_n\cdots z_k}(v'_0)$. Since $B_{z_{k+1}} (B^{\diamond}_{\cent z_1\cdots z_k}(v_0)) = B^{\diamond}_{\cent z_1\cdots z_{k+1}}(v_0) \subseteq \tilde{B}^{\diamond}_{\cent z_n\cdots z_k}(v'_0)$, if we apply $\tilde{B}_{z_{k+1}}$ ($ = B_{z_{k+1}}^{-1}$), then it follows that $B^{-1}_{z_{k+1}}(B^{\diamond}_{\cent z_1\cdots z_{k+1}}(v_0)) \subseteq B^{-1}_{z_{k+1}} (\tilde{B}^{\diamond}_{\cent z_n\cdots z_k}(v'_0))$.
Lemma \ref{inverse-basic}(1) further implies that $B^{\diamond}_{\cent z_1\cdots z_k}(v_0) \subseteq \tilde{B}^{\diamond}_{\cent z_1\cdots z_{k+1}}(v'_0)$, as requested.

Next, we target Statement (2). Assume that $v_0\in \tilde{B}^{\diamond}_{\cent z\dollar}(v_{acc})$. Since $\{v_0\}\cap \tilde{B}^{\diamond}_{\cent z\dollar}(v_{acc})\neq \setempty$, by taking $B=B_{\cent}$ in Lemma \ref{inverse-basic}(2), we obtain $B_{\cent}(v_0) \cap \tilde{B}^{\diamond}_{\cent z}(v_{acc}) \neq\setempty$.
By induction hypothesis, we assume that $B^{\diamond}_{\cent z_nz_{n-1}\cdots z_{k}}(v_0) \cap  \tilde{B}^{\diamond}_{\cent z_1z_2\cdots z_{k+1}}(v_{acc}) \neq\setempty$. Since $\tilde{B}^{\diamond}_{\cent z_1\cdots z_{k+1}}(V_{acc}) = B^{-1}_{z_{k+1}} (\tilde{B}^{\diamond}_{\cent z_1\cdots z_k}(v_{acc}))$, we apply  Lemma \ref{inverse-basic}(2) again with $B= B_{z_{k+1}}$. It then follows that  $B_{z_{k+1}} (B^{\diamond}_{\cent z_nz_{n-1}\cdots z_{k}}(v_0) ) \cap  \tilde{B}^{\diamond}_{\cent z_1z_2\cdots z_k}(v_{acc}) \neq\setempty$; in other words, $B^{\diamond}_{\cent z_nz_{n-1}\cdots z_{k+1}}(v_0) \cap  \tilde{B}^{\diamond}_{\cent z_1z_2\cdots z_k}(v_{acc}) \neq\setempty$.
Thus, by mathematical induction, Statement (2) is true.
\end{proof}

\subsection{Relationships between 1dta's and 1nta's}

In a general setting, \emph{nondeterminism} seems more powerful than determinism; however, it is known that 1nfa's can be simulated by appropriate 1dfa's at the cost of exponentially more inner states than the original 1nfa's. Here, we seek a direct simulation of 1nta's by appropriate 1dta's. In the following theorem, for a given topological space $(V,T_V)$, we expand $V$ to ${T^{+}_V}$ ($=T_V-\{\setempty\}$) so that $(T_V^{+},T^{\circ}(T_V^{+}))$ forms a topological space for an appropriately chosen topology $T^{\circ}(T^{+}_V)$.  Following Michael \cite{Ern51}, we here take $T^{\circ}(T^{+}_V)$ as the topology that is generated by the bases $\{[A]^{+},[A]^{-}\mid A\in T_V\}$, where $[A]^{+}=\{X\in {T^{+}_V} \mid X\subseteq A\}$ and $[A]^{-}=\{X\in {T^{+}_V} \mid X\cap A\neq\setempty\}$. This topology  is known as the
\emph{Vietoris topology},  adapted to ${T^{+}_V}$. Let us recall from Section \ref{sec:topology-notion} the notation $\co T_V$ for any topology $T_V$.

\begin{ytheorem}\label{Vietoris-topology}
Let $(\VV,\BB,\OO)$ be any extended automata base.  There exists
an automata base $(\VV',\BB',\OO')$ with
$\VV'=\{({T^{+}_V},T^{\circ}(T^{+}_V)) \mid V\in\VV\}$ such that, for any   $(\VV,\BB,\OO)$-1nta $M$ with $v_0$ and $V$, there is an equivalent $(\VV',\BB',\OO')$-1dta $N$, provided that $\{v_0\}\in T_V^{+}$.
\end{ytheorem}

\begin{proof}
From a given extended automata base $(\VV,\BB,\OO)$, since $\VV'$ is already given in the premise of the proposition, we only need to define the remaining $\BB'$ and $\OO'$. For each space $V\in\VV$, let us consider the set $T_V^{+}$.
Given a multi-valued operator $B: V\to V$ and an element $W\in T^{+}_V$, we define a single-valued operator $B':T_V^{+}\to T_V^{+}$ by setting $B'(W) = \bigcup_{w\in W}B(w)$.
Let $\BB'$ be composed of all sets $F'= \{B':{T^{+}_V}\to {T^{+}_V}\mid V\in \VV, B\in F\}$ for each $F\in\BB$. Finally, $\OO'$ consists of all pairs $(E'_1,E'_2)$ in ${T^{+}_V}\times {T^{+}_V}$ for any $V\in\VV$ such that (i)  $E'_1,E'_2\in T^{\circ}(T^{+}_V)\cap\co{T^{\circ}(T^{+}_V)}$ with  $E'_1\cap E'_2=\setempty$ and (ii) there exists an observable pair $(E_1,E_2)\in \OO$ satisfying both  $A_1\cap E_{1}\neq\setempty$ and $A_2\subseteq E_{2}$ for any $A_1\in E'_1$ and $A_2\in E'_2$.

Next, we argue that $(\VV',\BB',\OO')$ forms a valid automata base.
Given a set $V\in\VV$, take any subset $F$ of $C_m(V)$ in $\BB$.
Let $B$ be any  multi-valued continuous operator in $F$ and take its corresponding single-valued operator $B'$ in $F'$.
We wish to prove that $B'$ is continuous on the topological space $(T_V^{+},T^{\circ}(T_V^{+}))$. For this purpose, assume that $B'(W)=U$ holds for two arbitrary elements $U,W\in T_V^{+}$ and consider any open set $S$ in $T^{\circ}(T^{+}_V)$ containing $U$. Without loss of generality, we assume that $S$ is either $[U]^{-}$ or $[U]^{+}$ because $U$ is a nonempty open set of $V$. In the case of  $S=[U]^{-}$,  we set $R=[W]^{-}$. For any element $Y\in R$, it follows that $B'(Y)\subseteq U$, and thus $B'(Y)\in S$. In contrast, when $S=[U]^{+}$, we set $R=[W]^{+}$ instead.
Given any $Y\in R$, since $W\cap Y\neq\setempty$, we obtain $U\cap B'(Y)\neq\setempty$; hence, $B'(Y)\in S$ follows.

Let $M=(\Sigma,\{\cent,\dollar\},V, \{B_{\sigma}\}_{\sigma\in\check{\Sigma}}, v_0,E_{acc},E_{rej})$ be any $(\VV,\BB,\OO)$-1nta satisfying $\{v_0\}\in T_V^{+}$.
We define $E'_{acc}$ to be the set $\{E'\in T^{\circ}(T^{+}_V)\cap \co{T^{\circ}(T^{+}_V)} \mid \forall A\in E'[A\cap E_{acc}\neq\setempty]\}$ and  $E'_{rej}=\{E'\in T^{\circ}(T^{+}_V)\cap \co{T^{\circ}(T^{+}_V)} \mid \forall A\in E'[A\subseteq  E_{rej}]\}$.
Let us consider a 1dta $N$ of the form $(\Sigma,\{\cent,\dollar\},T^{+}_V, \{B'_{\sigma}\}_{\sigma\in\check{\Sigma}}, v'_0,E'_{acc},E'_{rej})$ with  $v'_0=\{v_0\}$. Clearly, $N$ is a $(\VV',\BB',\OO')$-1dta. To complete our proof, we need to verify that $N$ is computationally equivalent to $M$.

Toward our goal, we first prove that $B^{\diamond}_{w}(v_0) = B'_{w}(v'_0)$ holds for any extended input $w\in \{\cent\}\Sigma^* \{\dollar,\lambda\}$. In the base case of $w=\cent$, since  $B^{\diamond}_{\cent}(v_0)=B_{\cent}(v_0)$ and $B'_{\cent}(v'_0) = \bigcup_{w\in v'_0}B_{\cent}(w)=B_{\cent}(v_0)$, we conclude that  $B^{\diamond}_{\cent}(v_0)=B'_{\cent}(v'_0)$. For an induction step,  assume  that $B^{\diamond}_{\cent x}(v_0)=B'_{\cent x}(v'_0)$. Let us consider an extended input of the form $xa$ with $a\in\Sigma\cup\{\dollar\}$ and write  $U_{x}$ for $B^{\diamond}_{\cent x}(v_0)$. It follows that $B^{\diamond}_{\cent xa}(v_0) = (B_a\diamond B^{\diamond}_{\cent x})(v_0) =  B_a (B^{\diamond}_{\cent x}(v_0)) =  \bigcup_{w\in U_x}B_a(w)$ and that $B'_{\cent xa}(v'_0) = \bigcup_{w\in B'_{\cent x}(v'_0)} B_a(w) = \bigcup_{w\in U_x}B_a(w)$.
We then deduce that $B^{\diamond}_{\cent xa}(v_0)= B'_{\cent xa}(v'_0)$; in particular, $B^{\diamond}_{\cent x\dollar}(v_0) = B'_{\cent x\dollar}(v'_0)$ follows.

For all strings $x\in L(M)$, it follows that  $B^{\diamond}_{\cent x\dollar}(v_0) \cap E_{acc}\neq\setempty$ iff $B'_{\cent x\dollar}(v'_0)\cap E'_{acc}\neq\setempty$. On the contrary, if $x\notin L(M)$, then it follows that $B^{\diamond}_{\cent x\dollar} \subseteq E_{rej}$ iff $B'_{\cent x\dollar}(v'_0)$.
Therefore, $x$ is accepted  by $M$ iff $x$ is accepted by $N$. This concludes that $L(M)=L(N)$, as requested.
\end{proof}

\section{A Brief Discussion on Future Challenges}\label{sec:discussion}

In the past literature (e.g., \cite{Boz03,EK74,Jea07}), several mathematical models of topological automata were proposed and then studied on their own platforms, which are quite different from ours.  In order to categorize formal languages of various computational complexities, this paper has proposed new, general machine models of one-way deterministic and nondeterministic topological automata. The fundamental machinery of our new models is based on various choices of topologies ranging from the trivial topology to the discrete topology. Such topological automata are descriptionally powerful enough to represent  the existing finite automata of numerous types, including quantum finite automata,  pushdown automata, and interactive proof systems.

It turns out that topology and its associated concepts are quite expressible  to describe language families. In Section \ref{sec:new-model}, we have listed four key goals of the study of topological automata. Our study conducted in this paper is merely the initial step to fulfill these goals but it is still far away from the full understandings of the topological features that characterize various language families. To pave a road to a future study, we provide a short list of challenging open questions.

\begin{enumerate}\vs{-1}
  \setlength{\topsep}{-2mm}%
  \setlength{\itemsep}{1mm}%
  \setlength{\parskip}{0cm}%

\item The family $\reg$ of all regular languages is one of the most basic language families. We have given a few characterizations of $\reg$ in terms of topological automata, e.g., in Theorem \ref{iff-cond-uniform}. Find a more ``natural'' automata base $(\VV,\BB,\OO)$ that fulfills the equality of  $(\VV,\BB,\OO)\mbox{-}\mathrm{1DTA} = \reg$.

\item Complementing the first question, find ``natural'' automata bases $(\VV,\BB,\OO)$ and $(\VV',\BB',\OO')$ for which  $(\VV,\BB,\OO)\mbox{-}\mathrm{1DTA}\nsubseteq \reg$ and $\reg \nsubseteq (\VV',\BB',\OO')\mbox{-}\mathrm{1DTA}$.

\item In Proposition \ref{Vietoris-topology}, we have shown how to simulate each 1nta by a  computationally-equivalent 1dta.  Find a more ``succinct'' description of $(\VV',\BB',\OO')$-1dta that is computationally equivalent to any given  $(\VV,\BB,\OO)$-1nta.

\item The complexity classes $\dcfl$ and $\mathrm{MM\mbox{-}1QFA}$ are not closed under intersection.  Find a necessary and sufficient condition of  $(\VV,\BB,\OO)$ such that $(\VV,\BB,\OO) \mbox{-}\mathrm{1DTA}$ is not closed under intersection. This contrasts Lemma \ref{closure-basics}(3).

\item Given an automata base $(\VV,\BB,\OO)$ with ``natural'' topologies, characterize the language family $(\VV,\BB,\OO)\mbox{-}\mathrm{1DTA}$ in terms of standard automata.

\item In Section \ref{sec:normalization}, we have discussed a type of ``minimal'' topological automata. Find a ``natural'' notion of minimality for our models of topological automata and give an exact condition on $(\VV,\BB,\OO)$ that guarantees the existence of such  minimal $(\VV,\BB,\OO)$-1dta's.

\item We have discussed the Kolmogorov separation axiom in Section \ref{sec:compare-properties}. When an automata base $(\VV,\BB,\OO)$ violates the Kolmogorov separation axiom, what is the language family $(\VV,\BB,\OO)\mbox{-}\mathrm{1DTA}$?

\item Neither \emph{vector spaces} nor \emph{metric spaces} have been discussed in this paper although our framework of 1dta's is powerful enough to capture all languages. However, certain types of finite automata are originally defined on those spaces. For example, quantum finite automata are founded on Hilbert spaces with the $\ell_2$-norm. Develop a coherent theory of topological automata that are based on vector spaces or metric spaces.

\item In this paper, we have discussed only the case where any computation evolves \emph{in linear fashion}. If we further expand our basic models using \emph{nonlinear evolutions}, how do the corresponding one-way finite automata look like?
\end{enumerate}



\let\oldbibliography\thebibliography
\renewcommand{\thebibliography}[1]{%
  \oldbibliography{#1}%
  \setlength{\itemsep}{0pt}%
}
\bibliographystyle{plain}

\end{document}